\documentclass{emulateapj}

\usepackage{graphicx}
\usepackage{enumerate}
\usepackage{amsmath}
\usepackage{comment}

\def\be{\begin{equation}}
\def\ee{\end{equation}}
\def\bea{\begin{eqnarray}}
\def\eea{\end{eqnarray}}
\def\bn{\begin{enumerate}}
\def\en{\end{enumerate}}

\def\sec\ond{{\rm s}}

\def\Mpc{{\rm Mpc}}

\def\hMpc{\,h^{-1}\Mpc}

\def\rms{{\frenchspacing r.m.s.}}
\def\be{\begin{equation}}\def\bea{\begin{eqnarray}}\def\beaa{\begin{eqnarray*}}
\def\ee{\end{equation}}  \def\eea{\end{eqnarray}}  \def\eeaa{\end{eqnarray*}}
\newcommand{\hmpc}{{h^{-1}\,{\rm Mpc}}}

\shorttitle{Topology of SDSS LRGs}
\shortauthors{Choi et al. (2013)}
\begin{document}
\title{Topology of Luminous Red Galaxies from the Sloan Digital Sky Survey}
\author{Yun-Young Choi$^1$, Juhan Kim$^{2,3}$, Graziano Rossi$^{4,5}$, Sungsoo S. Kim$^6$, \& Jeong-Eun Lee$^6$}
\affil{$^1$Department of Astronomy and Space Science, Kyung Hee University, Gyeonggi 446-701, Korea; yy.choi@khu.ac.kr\\
$^2$Center for Advanced Computation, Korea Institute for Advanced Study, Heogiro 85, Seoul 130-722, Korea\\
$^3$School of Physics, Korea Institute for Advanced Study, Heogiro 85, Seoul 130-722, Korea\\
$^4$CEA, Centre de Saclay, Irfu/SPP, F-91191 Gif-sur-Yvette, France \\
$^5$ Paris Center for Cosmological Physics and Laboratoire APC, Universit\'e Paris 7, 75205 Paris, France \\
$^6$School of Space Research, Kyung Hee University,
Gyeonggi 446-701, Korea}
\email{Corresponding Author: Graziano Rossi (graziano.rossi@cea.fr)}



\begin{abstract}

We present measurements of the genus topology of luminous red galaxies (LRGs) 
from the Sloan Digital Sky Survey (SDSS) Data Release 7
catalog, with unprecedented statistical significance. 
To estimate the uncertainties in the measured genus, we construct 81 mock 
SDSS LRG surveys along the past light cone
from the Horizon Run 3, one of the largest $N$-body simulations to date that 
evolved $7210^3$ particles in a $10815~h^{-1}$Mpc size box.
After carefully modeling and removing all known systematic effects due to 
finite pixel size, survey boundary, radial and angular selection functions,
shot noise and galaxy biasing, we find the observed genus amplitude 
to reach 272 at 22~$h^{-1}$Mpc smoothing scale with an uncertainty of 4.2\%;
the estimated error fully incorporates cosmic variance.
This is the most accurate constraint of the genus amplitude to date, 
which significantly improves on our previous results.
In particular, the shape of the genus curve agrees very well with
the mean topology of the SDSS LRG mock surveys in the $\Lambda$CDM universe.
However, comparison with simulations also shows small deviations of the 
observed genus curve from the theoretical expectation for Gaussian 
initial conditions. 
While these discrepancies are mainly driven by known systematic effects
such as those of shot noise and redshift-space distortions,
they do contain important cosmological information on the physical effects
connected with galaxy formation, gravitational evolution and primordial 
non-Gaussianity. 
We address here the key role played by systematics on the genus curve, 
and show how to accurately correct for their effects 
to recover the topology of the underlying matter.
In a forthcoming paper, we provide an interpretation of  those deviations 
in the context of the local model of non-Gaussianity.

\end{abstract}






\section{Introduction}


The current standard cosmological scenario, supported by observations of the 
cosmic microwave background (CMB)
and of the large-scale structure (LSS), appears to be consistent with the 
$\Lambda$CDM concordance model,
where the Universe is dominated by cold dark matter (CDM) and 
its accelerating expansion driven by a cosmological constant $\Lambda$ 
or dark energy (DE).
A recent strong support of this paradigm has been presented by 
Park et al. (2012), who was able to prove that 
observed high- and low-density LSSs have the richness/volume 
and size distributions consistent with the $\Lambda$CDM universe. 

In addition, the primordial density perturbations from which halos and 
galaxies form are assumed to be a Gaussian random field, as predicted by 
inflationary theories (Guth 1981; Linde 1982; Bardeen et al. 1986); 
state-of-the-art data from the  Wilkinson Microwave Anisotropy Probe 
(WMAP; Spergel et al. 2003, Komatsu et al. 2011), 
the Sloan Digital Sky Survey (SDSS; York et al. 2000; Stoughton et al. 2002;
Abazajian et al. 2009) 
or the WiggleZ survey (Blake et al. 2011) are still favoring this hypothesis.
However, some claims or hints of primordial non-Gaussianity have recently 
appeared in the literature 
(Jeong \& Smoot 2007; Yadav \& Wandelt 2008; Komatsu et al. 2009, 2011; 
Slosar et al. 2008; Smidt et al. 2010),
and challenged the validity of the simplest inflationary paradigm.
Indeed, if detected, primordial non-Gaussianity would indicate a 
structure formation scenario different from the concordance cosmological 
model, and force us to revise the physics of the very early 
Universe -- along with several aspects of the LSS dynamics
(but see also Hwang 2012 for a more general discussion on modern cosmology).


To this end, topology-related statistics offer a precious benchmark for 
testing the underlying Gaussianity of the initial density field, 
since topology can be regarded as an important physical property of the 
matter density that can be compared with predictions of
the simplest inflationary models, 
where Gaussian random phase initial conditions are generated from quantum 
fluctuations of an inflaton field in the early Universe. 
In addition,  topology measured at the present epoch 
should reflect that of the initial conditions on smoothing scales 
considerably larger than the correlation length, because
fluctuations which are still in linear regime maintain their initial 
topology (see Gott et al. 1987,  who confirmed this property
with $N$-body simulations); this fact allows one to test directly 
the Gaussian paradigm, and permits to use topology as a cosmic standard 
ruler (Park \& Kim 2010).


From the theoretical side, since the pioneering work of Gott et al. (1986),
a variety of analytic and numerical tools to analyze 
observational and simulated data for measuring topology have long been 
developed -- mainly using the genus statistics to quantify the topology of 
isodensity contours (Gott et al. 1987, 1989; Hamilton et al. 1986; 
Vogeley et al. 1994; Park et al. 2005a, 2005b).
In particular,  
the analytic prediction for the genus curve of a Gaussian field
in linear regime is well-known (Gott et al. 1986), and   
its perturbative expression in the weakly nonlinear regime 
has also been obtained (Matsubara 1994);
this lognormal model  turned out to be a good empirical 
approximation in the strongly nonlinear regime (Hikage et al. 2002).
Along with analytic tools, large-volume $N$-body simulations are routinely 
used to quantify several systematics which affect the genus curve, such 
as finite pixel size, sparse sampling, peculiar velocity distortions in 
redshift space or survey boundaries.
The ability to correct for these effects is essential, as the remaining 
small deviations from the random phase curve give important
information about the physics connected with galaxy formation, nonlinear 
gravitational clustering, and primordial non-Gaussianity -- if any.
In fact, on smaller scales nonlinear gravitational evolution and biased 
galaxy formation make the topology of the observed galaxy distribution 
deviate from the Gaussian form, even if the initial conditions
were Gaussian-distributed.
Using fractional volume rather than direct density 
threshold as the independent variable in topology analysis mitigates 
but does not fully eliminate these nonlinear and biasing effects 
(Weinberg et al. 1987; Melott et al. 1988).
Ultimately, all the secondary non-Gaussianities need to be
disentangled from the primordial contribution, and this can now be done very accurately, without assuming any `a priori' model for the underlying signal.


From the observational side, a long list of studies have been pursued 
on a variety of datasets (see for example Park, Gott, \& da Costa 1992;
Park et al. 1992, 2005b; Park, Gott, \& Choi 2001; 
Moore et al. 1992; Vogeley et al. 1994; Rhoads et al. 1994;
Protogeros \& Weinberg 1997; Canavezes et al. 1998; Hoyle et al. 2002; 
Hikage et al. 2002, 2003; James et al. 2007, 2009; Gott et al 2008, 2009; 
Choi et al. 2010).
They focused on characterizing the three-dimensional topology, and
showed that, depending on the considered scale, topology is
useful in constraining both cosmological parameters and the galaxy 
formation mechanism.
For example, 
Park et al. (2005b)  characterized the topology of the SDSS 
Main galaxy sample 
from the NYU Value Added Galaxy Catalog (NYU VAGC; Blanton et al. 2005), which has 
similar sky coverage as the SDSS Data Release 4 
(DR4; Adelman-McCarthy et al. 2006),
and presented the first clear demonstration of luminosity dependence of 
galaxy clustering topology (i.e. brighter galaxies show a stronger signal of 
meatball topology); more recently, Gott et al. (2009) measured the three-dimensional LSS 
topology of the SDSS DR4plus LRG sample from the NYU VAGC
(a subsample of the SDSS DR5; Adelman-McCarthy et al. 2007)
and found strong consistency 
with Gaussianity of the primordial fluctuations.
In the latter case,  the large sample size available allowed topology to 
be an important tool for testing galaxy formation models.
Also, Choi et al. (2010)  measured the topology of the Main galaxy 
distribution using the SDSS DR7 (Abazajian et al. 2009; Choi, Han, \& Kim 2010), studied the 
scale-dependent topology bias, and examined the dependence of galaxy 
clustering topology on galaxy properties (i.e. luminosity, morphology, color) 
at different smoothing scales.
The large volume-limited sample enabled an unprecedented measurement of the 
genus curve, with an amplitude of $G=378$ at 6 $h^{-1}$Mpc smoothing scale 
and an estimated uncertainty of 4.8$\%$, 
including all systematics and cosmic variance.
In addition, Choi et al. (2010) detected deviations of the genus curve 
from Gaussianity, interpreted as the fact that voids and superclusters 
are more connected and their sizes are larger than those expected for 
Gaussian fields.
The same authors then used these results to test five different galaxy 
formation models, which indeed are tuned to reproduce the 
two-point correlation function and the luminosity function but not 
high-order statistics, and found significant discrepancies: none of 
the models could reproduce all the aspects of the observed clustering topology.


While the significant discrepancies in the SDSS DR7 galaxy clustering 
topology at nonlinear scales found by Choi et al. (2010) are mainly 
driven by the inaccuracy of galaxy formation models, a cleaner problem 
is to consider the topology of LRGs instead -- which is the main focus of 
this paper. 
This is because the LRG sample covers a much larger and deeper volume 
(i.e. it allows one to observe topology at the largest scales), 
essentially in the linear regime.
In addition, LRGs are particularly useful in refining cosmological 
parameters (Tegmark et al. 2006), and are expected to play an important 
role in characterizing DE through the ratio of DE pressure to energy density 
(see Bassett et al. 2005; Eisenstein et al. 2005; Percival et al. 2007). 
A number of previous analyses have addressed the clustering of LRGs, 
especially in relation to Baryon Acoustic Oscillation (BAO) science 
(see for instance Ross et al. 2012, and references therein).
On the contrary, fewer studies have been based on topology. 
Among those, Gott et al. (2009) measured the three-dimensional genus 
topology of LRGs, using two volume-limited samples constructed from 
the SDSS DR4plus sample,
i.e. a dense shallow sample with $21\hMpc$ smoothing, and a 
sparse deep sample with $36\hMpc$ smoothing. 
The amplitude of the genus curve  was found to reach about 167
with a $4.1\%$ uncertainty at $21\hMpc$ scale. 
A major result of their study was that topology of LRGs in the SDSS 
agrees very well with that of mock galaxies in the $\Lambda$CDM universe with 
the same cosmological parameters:
small distortions in the genus curve, expected from nonlinear biasing and 
gravitational effects, are well explained by $N$-body simulations
with a subhalo finding technique adopted to locate LRGs. 
This suggests that the formation of LRGs  can be modeled well 
without any free-fitting parameters.


\begin{figure*}
\epsscale{0.9}
\plotone{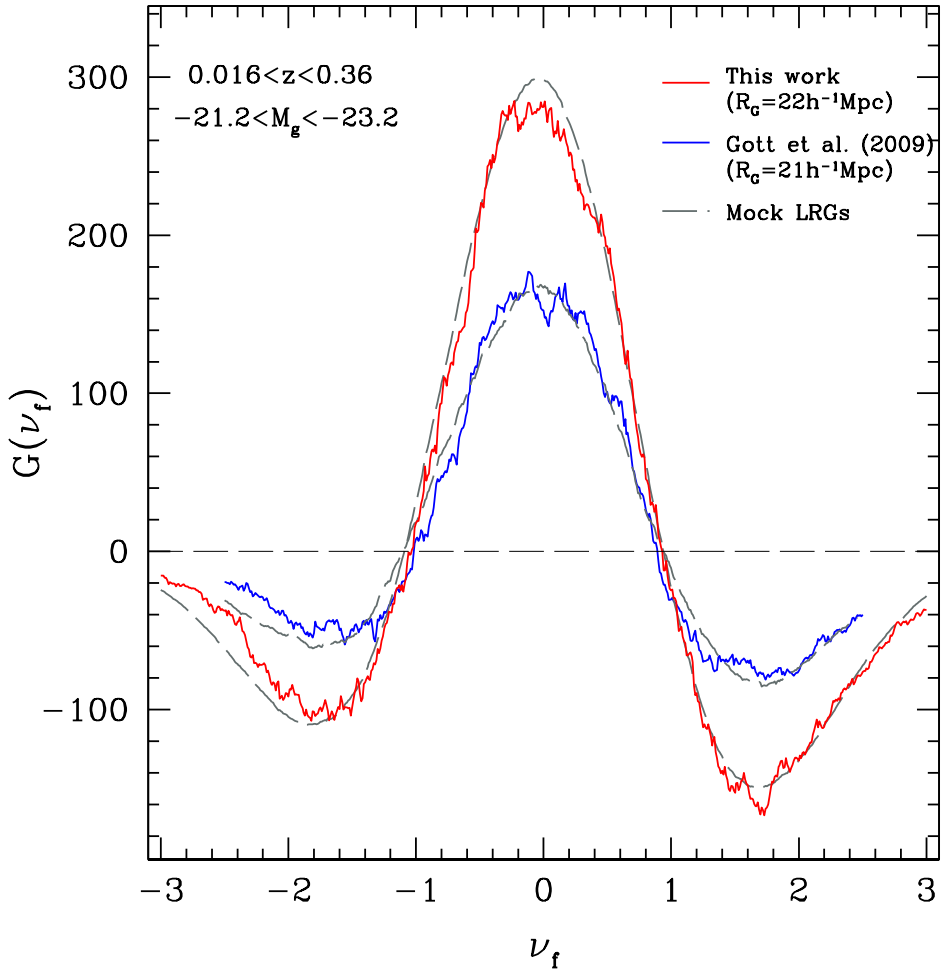}
\caption{Genus curve measured from the number density field of
the SDSS DR7 volume-limited sample (red solid line).
For illustrative purposes, the curve is compared to the one obtained 
by Gott et al. (2009) from the SDSS DR4plus SHALLOW sample 
(i.e. blue solid line).
The main point of the figure is to show the dramatic increase in the genus 
amplitude over the two different datasets, due to the much larger volume 
spanned by the SDSS DR7.
The observed genus curves in this plot have not been corrected 
for any systematic bias, since the figure is used just as an illustrative 
example.
In addition, while the two LRG samples considered are defined in the same way, 
slightly different smoothing lengths and pixel sizes have been applied 
to their corresponding density fields
(i.e. $21\hMpc$ and $7\hMpc$ for the DR4plus SHALLOW sample, 
and $22\hMpc$ and $3.7\hMpc$ for the DR7, respectively).
Grey dashed lines are genus curves averaged over
mock SDSS surveys in the simulated $\Lambda$CDM universe.
We will return to this measurement in great detail in 
Section~\ref{sec:genus_top}.}
\label{fig:main_result}
\end{figure*}


The main goal of this paper is to characterize the three-dimensional genus 
topology of spectroscopic LRGs using the SDSS DR7 catalog, 
improving on the previous results presented by Gott et al. (2009).
In particular, we strive to carefully model and remove all known 
systematics which affect the observed genus (i.e. finite pixel size, survey 
boundary, radial and angular selection function
and shot noise), and estimate the uncertainties in the measured genus 
accurately. 
This is achieved by comparing our measurements with 81 mock SDSS LRG surveys 
along the past light cone constructed from the Horizon Run 3 (HR3; Kim et al. 2011), 
one of the largest $N$-body simulations to date that 
evolved $7210^3$ particles in a $10815~h^{-1}$Mpc size box.
Our main result for the observed genus curve is shown in 
Figure~\ref{fig:main_result} (red solid line),
and compared with a previous topology measurement
on the SDSS DR4plus dataset (blue solid line; Gott et al. 2009).
The main point of the figure is to show the dramatic increase in amplitude 
of the genus curve over the two different datasets (i.e. the SDSS DR4plus
versus the SDSS DR7) in the same redshift range from 0.16 to 0.36 and 
rest-frame $\rm g$-band absolute magnitudes of $-23.2<M_{\rm g}<-21.1$,
due to the much larger volume that is covered.
In fact, we find the genus amplitude to reach 285 with an uncertainty 
of 4.0\% at $22\hMpc$ Gaussian smoothing scale 
including cosmic variance (the most accurate measurement to date), 
while Gott et al. (2009) found the genus curve to reach about 167
with a $4.1\%$ uncertainty at $21\hMpc$ smoothing scale;
for comparison with a different galaxy population, 
Park et al. (2005b) and Choi et al. (2010) reported an uncertainty of 
$9.4\%$ and $4.8\%$ at 5$h^{-1}$Mpc and 6$h^{-1}$Mpc scales, respectively,
for the genus obtained from the Main galaxy sample of the SDSS DR3 and DR7. 
We will return to this measurement in great detail in 
Section~\ref{sec:genus_top}, while in a forthcoming publication 
we interpret the deviation of the genus curve from  the expected Gaussian 
prediction in the context of primordial non-Gaussianity. 


The layout of the paper is organized as follows. 
In Section \ref{sec:theory}, we present the main theoretical framework of 
this study; 
in particular, we review the genus statistics for Gaussian fields 
and discuss how to extend the formalism for non-Gaussian fields. 
In Section~\ref{sec:data_methods}, we describe the SDSS 
LRG sample used for our measurements and the methodology applied to the 
observational data.
In Section \ref{sec:sims_methods}, we present the HR3 $N$-body simulation
and explain the procedure adopted to construct the 81 mock LRG surveys.
In Section \ref{sec:genus_top}, we
show our results  for the LRG genus statistics, compare measurements from
SDSS DR7 data and simulations and quantify the non-Gaussian deviations of
the genus curve with genus-related quantities.
In Section~\ref{sec:genus_sys}, we discuss the effects of known systematics on
the genus, and present the genus curve after corrections for systematics.
We conclude in Section~\ref{sec:summary}, and leave some more details 
on the genus curves in the Appendix.



\section{Theoretical Background} \label{sec:theory}

We begin by revisiting the basic theory of the genus 
for Gaussian fields and by introducing some genus-related statistics.
We also briefly review the formalism for describing the non-Gaussian effect
on the genus curve in the weakly nonlinear regime
according to second-order perturbation theory, originally derived by
Matsubara (1994, 2003).
We will later compare the theory outlined here with results from the SDSS 
DR7 LRG dataset, and with measurements from simulated LRG sample.
The full extension to non-Gaussian fields with the inclusion of primordial
non-Gaussianity will be presented and discussed in a forthcoming publication.


\subsection{Genus statistics for Gaussian random fields}

The genus is a measurement of the topology of isodensity contour surfaces in a smoothed galaxy field (Gott et al. 1987).
In mathematical terms, it is defined as follows. 
Consider a three-dimensional Gaussian random field $\rho \equiv \rho({\bf x})$ 
with ${\bf x}$ the spatial coordinate, and measure the topology of the 
excursion regions where $\rho$ is equal to, or is above a given threshold 
level $\bar{\rho} + \nu \sigma_0$. Here $\bar{\rho}$ 
is the mean of the field $\rho$,
$\sigma_0$ its root mean square ($\rms$) value, 
and $\nu=(\rho - \bar{\rho})/\sigma_0$. Denote with $M$
the space which contains the set of the excursion regions (i.e. a 3-manifold 
subset), and indicate its boundaries with $\delta M$ (i.e. a 2-manifold 
subset). 
For each component $S_{\rm i}$ of $\delta M$, 
according to the Gauss-Bonnet theorem, the mathematical
genus satisfies the following relation
\begin{equation}
{\cal G}_{\rm i} = I - \frac{1}{2} \chi(S_{\rm i}),
\end{equation}
where $\chi(S_{\rm i})$ is the Euler characteristic of the surface of the
three-dimensional excursion region (i.e. the integrated
Gaussian curvature of the surface).
Hence, the total genus of the boundary $\delta M$ becomes
\begin{equation}
{\cal G} = \sum_{\rm i} {\cal G}_{\rm i} = N - \frac{1}{2} \sum_{\rm i} \chi(S_{\rm i}) = N -\frac{1}{2} \chi(\delta M),
\label{formula_genus}
\end{equation}
where $N$ is the number of components of $\delta M$ -- see Park et al. (2013) for a full derivation of the previous formula. 

In cosmology, the standard definition of genus slightly differs from the previous mathematical one, since the genus is defined as
the number of holes minus the number of isolated regions
in the isodensity contour surfaces, at a given threshold level $\nu$. Namely,
\begin{eqnarray}
G(\nu) &=& {\rm Number~of~holes~in~contour~surfaces} - \nonumber\\
&&{\rm Number~of~isolated~regions}.
\end{eqnarray}
The relation between the two definitions is simply expressed by $G = {\cal G}- N$, where $N$ has been defined above.
For further insights on topological invariants, and 
for the mathematical  connection between the cosmological genus and the Betti numbers for excursion sets of Gaussian random fields,
we refer the reader to Park et al. (2013). 

In the case of Gaussian fields, the genus per unit volume $g(\nu)=G(\nu)/V$ as a
function of density threshold level $\nu$ is known (i.e. Doroshkevich 1970; Adler 1981;
Hamilton, Gott, \& Weinberg 1986):
\begin{equation}
g(\nu)=g(0) (1-\nu^{2}) {\rm exp} ({-\nu^{2}/2}).
\label{eq_genus_standard}
\end{equation}
The amplitude $g(0)$ is given by
\begin{equation}
g(0) = {1\over {(2\pi)^2}} \Big ({\sigma_1 \over {\sqrt{3}\sigma_0}} \Big )^3,
\label{eq_genus_amplitude}
\end{equation}
while the spectral moments of the fields, $\sigma_{\rm j}^2$, are computed from
\begin{equation}
\sigma_{\rm j}^2 (R_{\rm G}) = {1 \over (2 \pi)^2} \int k^{\rm 2(j+1)}P(k,z) {\rm d k}. \label{eq:sigma}
\end{equation}
In the previous relation, $P(k,z)$ is the power spectrum smoothed on a scale
$R_{\rm G}$ by a window function $W$, where
\begin{equation}
P(k,z)=P_{\rm m}(k,z)  W^2(kR_{\rm G})
\end{equation}
and $P_{\rm m}(k,z)$ is the matter power spectrum.
In particular, in this study we adopt a Gaussian smoothing of the form
$W(kR_{\rm G})= {\rm exp}(-k^2 R_{\rm G}^2/2)$.
Note also that, for $j=0$,
$\sigma_{\rm j}$ is the variance of the fluctuating field, while
$\sigma_{\rm j}$ is the variance of its derivative when $j=1$.

To separate the
variation in topology from the change of the one-point density
distribution, in this work we also measure the genus as a function of the volume-fraction
threshold $\nu_{\rm f}$ (as opposed to the direct density threshold $\nu$).
This parameter defines the density contour surface
such that the volume fraction $f$ in the high density region is the same
as the volume fraction in a Gaussian random field contour surface having
$\nu = \nu_{\rm f}$, namely:
\begin{equation}
 f = {1 \over \sqrt{2 \pi} } \int_{\nu \equiv \nu_{\rm f}}^{\infty} {\rm exp}(-x^2/2) {\rm d}x.
\end{equation}


\subsection{Genus statistics and perturbation theory}

After correcting for known systematics, deviations of the observed genus curve
from the Gaussian expectation (i.e. Eq.~\ref{eq_genus_standard})
are due to nonlinear gravitational
evolution and non-Gaussianity of the primordial density field.
A number of studies in the literature have already addressed the impact of
non-Gaussianity on the genus curve (see for example Weinberg et al.
1987; Park \& Gott 1991; Park, Kim, \& Gott 2005a).
In what follows, we briefly discuss the non-Gaussian effect on the genus curve 
caused by nonlinear gravitational
evolution in the weakly nonlinear regime (which tends to distort the Gaussian 
expectation for the genus statistic), in the context of  second-order 
perturbation theory -- along the lines of Matsubara (1994, 2003).
More details on the non-Gaussian modifications of the genus curve will be
presented in Young-Rae Kim et al. (in preparation).  

To first order in $\sigma_0$, 
the nonlinear correction that one must apply to the genus curve due to 
gravitational evolution is an odd function of the threshold $\nu$. 
Hence, this correction causes a shift and an asymmetry
between high- and low-density regions, with no change in the amplitude at $\nu=0$.

In particular, when we use a threshold rescaled by a volume fraction
of the smoothed field, $\nu_{\rm f}$, 
the genus of the matter density field per unit volume -- expanded to first 
order in mass variance $\sigma_0$ -- can be
written as a sum of a Gaussian term $g^{\rm G}(\nu_{\rm f})$ 
plus a non-Gaussian term
$g^{\rm NG}(\nu_{\rm f})$, namely:
\begin{equation}
g(\nu_{\rm f})=g^{\rm G}(\nu_{\rm f})+g^{\rm NG}(\nu_{\rm f}).
\end{equation}
The Gaussian part is expressed by
\begin{equation}
g^{\rm G}(\nu_{\rm f})=-g(0)~{\rm exp}({-\nu_{\rm f}^2 /2})~H_2(\nu_{\rm f}),
\end{equation}
while the non-Gaussian term is given by (Young-Rae Kim et al., in preparation):
\begin{eqnarray}
\label{eq:matsubara}
&g&^{\rm{NG}}(\nu_{\rm f})=-g(0)~{\rm exp}(-\nu_{\rm f}^2 /2)\times\\
&&\left[(S^{(1)}-S^{(0)})H_3(\nu_{\rm f})+
(S^{(2)}-S^{(0)})H_1(\nu_{\rm f})\right]\sigma_{0}. \nonumber
\end{eqnarray}
In the previous equations,
$H_{\rm n}(\nu_{\rm f})$ are Hermite polynomials, and in particular $H_0(\nu_{\rm f})=1$,
$H_1(\nu_{\rm f})=\nu_{\rm f}$, $H_2(\nu_{\rm f})=\nu_{\rm f}^2-1$, $H_3(\nu_{\rm f})=\nu_{\rm f}^3-3\nu_{\rm f}$,
$H_4(\nu_{\rm f})=\nu_{\rm f}^4-6\nu_{\rm f}^2+3$, and $H_5(\nu_{\rm f})=\nu_{\rm f}^5-10\nu_{\rm f}^3+15\nu_{\rm f}$.
Also, the various $S^{\rm (a)}$, $a=0,1,2$,
are skewness parameters obtained by integrating the bispectrum
$B(k_1,k_2,k_3,z)$ over $\bf k_1$ and $\bf k_2$
(see Eq. 61-64 in Matsubara 2003).
In addition, the
bispectrum can be given in terms of the nonlinear contributions
from nonlinear gravitational evolution, and also primordial
non-Gaussianity -- an aspect that we do not consider here (but see Appendix B in Hikage et al.
2006).
Note that the non-Gaussian part $g^{\rm NG}(\nu_{\rm f})$ only appears with terms
of the form $S^{\rm (a)} -S^{\rm (0)}$.
Assuming galaxy biasing local and deterministic in the weakly nonlinear regime,
the skewness parameters of the galaxy bispectrum, $S_{\rm g}^{(\rm a)}$, is given by
$S_{\rm g}^{(a)}={S^{(a)}/b}+{{3b_2}/b^2},\label{Eq:bias}$
where $b$ and $b_2$ are bias parameters.
The non-Gaussian term
$g^{\rm NG}(\nu_{\rm f})$ in Equation~\ref{eq:matsubara} 
only appears as combinations of the form 
$(S_{\rm g}^{(a)}-S_{\rm g}^{(0)})\sigma_{\rm 0,g}$
when the biased variance at first order is then
given by $\sigma_{\rm 0,g}=b\sigma_0$; 
therefore, the non-Gaussian correction of the
galaxy density field is exactly the same as the one of the unbiased mass density
field in Equation~\ref{eq:matsubara} -- 
hence independent of the bias parameter.
This can be considered as an advantage of the volume fraction threshold, as opposed to the
direct density threshold (see Matsubara 2003 for more details).


\subsection{Genus-related statistics} \label{sec:grs}

The measured genus curve can be compared with predictions of
the simplest inflationary models, which assume Gaussian random phase initial conditions (and so with Eq.~\ref{eq_genus_standard}). 
However, even if the initial conditions were perfectly Gaussian,
small deviations from Gaussianity are expected because of systematic effects (for example shot noise or redshift-space distortions), and because
of  the physics connected with galaxy formation, nonlinear gravitational evolution, and primordial non-Gaussianity (if any).
Therefore, it is important to quantify even small departures from Gaussianity
from the observed genus. This is done by parametrizing the genus curve 
with several derived quantities. In what follows, we consider
measurements as a function of the volume fraction threshold $\nu_{\rm f}$, and introduce four genus-related statistics.
The first quantity is simply the best-fit genus amplitude 
$G_{\rm fit}(0)$,
measured by a least-squares fit of the theoretical random phase curve to the data considering only the range  $-1\leq\nu_{\rm f}\leq 1$.
In principle, its value is given by Equation~\ref{eq_genus_amplitude}, 
but the measured one is
always lower because of nonlinear clustering and biasing due
to coalescence of structures (Park \& Gott 1991; Vogeley et al.
1994; Canavezes et al. 1998; Gott et al. 2008).

The second quantity  is the shift parameter $\Delta\nu_{\rm f}$, defined as
\begin{equation}
\Delta\nu_{\rm f} = \int_{-1}^{1} G_{\rm o}(\nu_{\rm f})\nu_{\rm f}  {\rm d}\nu_{\rm f}   \Big / \int_{-1}^{1} G_{\rm fit}(\nu_{\rm f})  {\rm d} \nu_{\rm f},
\end{equation}
where $G_{\rm o}$
and $G_{\rm fit}$ are the observed and best-fit Gaussian genus curves -- both given by Equation \ref{eq_genus_standard},
but in the former case with the observed amplitude $G_{\rm o}(0)$ and in the latter with the best-fit one, $G_{\rm fit}(0)$, as explained in
Park et al. (1992). The parameter $\Delta\nu_{\rm f}$
controls the horizontal shifts of the central part of the
genus curve.
For a density field dominated by voids, $\Delta\nu_{\rm f}$ is positive and we say
that the density field has a ``bubble-like'' topology.
For a cluster-dominated field, $\Delta\nu_{\rm f}$ is negative and we say that
the field has a ``meatball-like'' topology.

We then further introduce two additional quantities,
$A_{\rm C}$  and $A_{\rm V}$, which measure the
abundances of clusters (C) and voids (V), respectively, relative to the expectations for a Gaussian
random field. They are defined by the following relation
\begin{equation}
A=\int G_{\rm o}(\nu_{\rm f})  {\rm d}\nu_{\rm f} \Big / \int  G_{\rm fit}(\nu_{\rm f}) {\rm d}\nu_{\rm f} ,
\end{equation}
where the integration intervals are $+1.2 <\nu_{\rm f}< +2.2$ for $A_{\rm
C}$, and $-2.2 <\nu_{\rm f}< -1.2$ for $A_{\rm V}$ (Park, Gott, \& Choi 2001;
Park, Kim, \& Gott 2005a). These intervals are centered near the minima
of the Gaussian genus curve (i.e. $\nu_{\rm f}=\pm\sqrt{3}$), far away from the
thresholds where the genus is often affected by the shift
phenomenon.  The previous intervals also exclude extreme thresholds, where for
low density regions $\nu_{\rm f}$ is very sensitive to the exact density value. 
These parameters are defined so that
 the condition $A_{\rm C}>1$ ($A_{\rm V}>1$)  implies that more independent clusters (voids) are
observed, with respect to those predicted by a Gaussian field at a fixed volume fraction.
On the opposite, $A_{\rm C}<1$ ($A_{\rm V}<1$) means that fewer independent clusters (voids)
are seen.

The effects of gravitational
evolution, galaxy biasing, and cosmology dependence on
the statistics defined by $\Delta\nu_{\rm f}$,
$A_{\rm V}$, and $A_{\rm C}$ -- as a function of the smoothing scale --
have been addressed in detail
 by Park, Kim, \& Gott (2005a); we refer to their study for more details.




\section{Observed LRG sample: Description and Methodology}  
\label{sec:data_methods}

In this section we briefly describe our 
LRG sample obtained from the SDSS DR7, 
along with the methodology applied to the observational dataset.
The genus computed from the LRG sample and its related statistics
will be presented later on, in Section~\ref{sec:genus_top}.


\subsection{The SDSS DR7 LRG sample}

\begin{figure*}
\epsscale{1.2}
\plotone{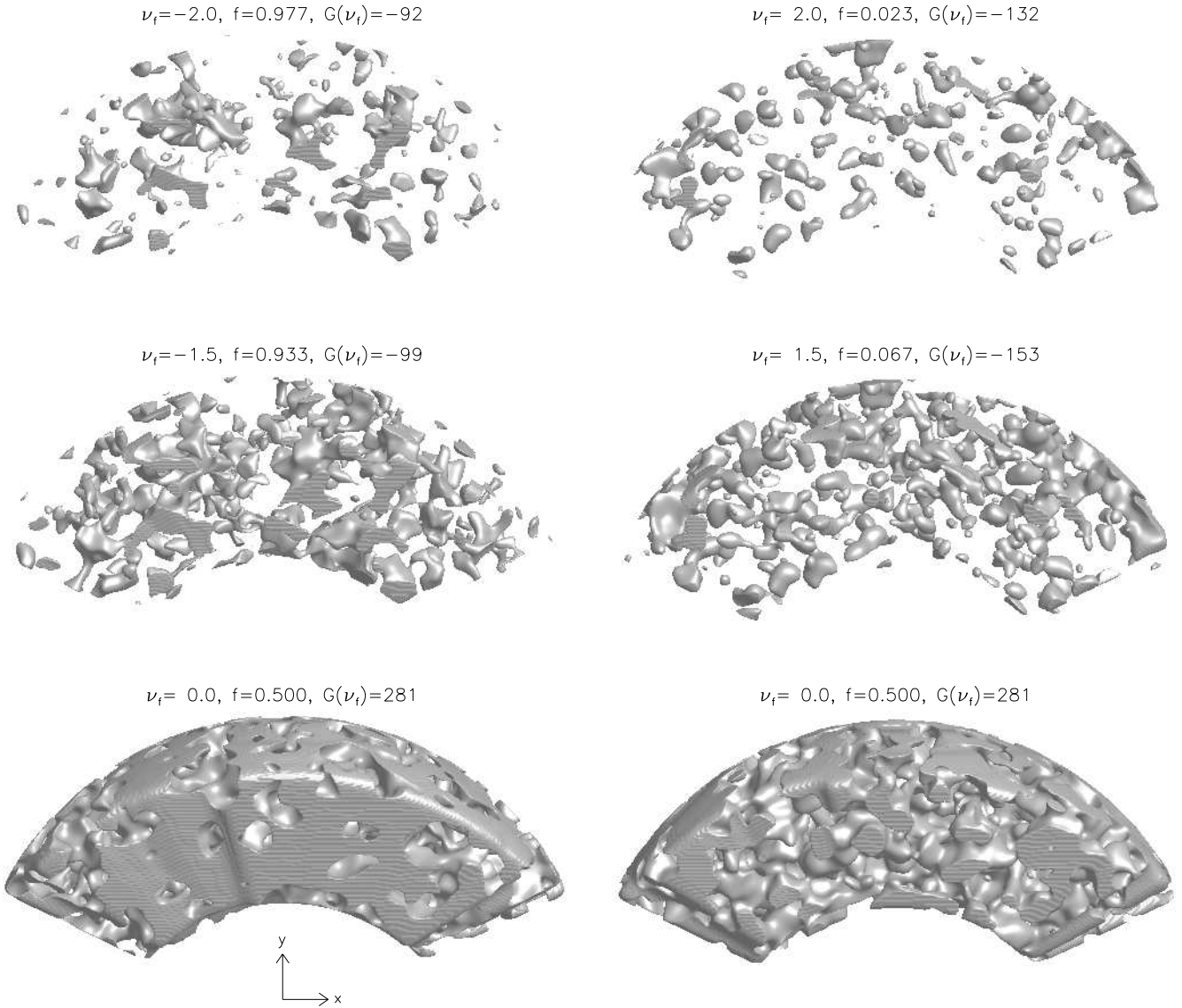}
\caption{
Three-dimensional view of the galaxy number density field from the SDSS
LRG volume-limited sample, smoothed with a Gaussian filter 
at $R_{\rm G} = 22\hMpc$ scale.
({\it Left}) From top to bottom, three representative 
density contours enclosing low-density
regions, which occupy respectively 2.3\% ($\nu_{\rm f}=-2.0$), 
6.7\% ($\nu_{\rm f}=-1.5$),
and 50\% ($\nu_{\rm f}=0.0$) of the sample volume.
({\it Right}) Three density contours enclosing now high-density
regions, filling respectively 
2.3\% ($\nu_{\rm f}=2.0$), 6.7\% ($\nu_{\rm f}=1.5$), 
and 50\% ($\nu_{\rm f}=0.0$) of the sample volume -- from top to bottom,
in symmetry with the low-density cases.
The Earth is located at the center of the $x$-$y$ plane shown
in the figure, and the size of each axis corresponds to
a scale of $200\hMpc$.}
\label{fig:isosurface}
\end{figure*}

The SDSS is a successful ground-based survey, designed to 
explore the large-scale distribution of galaxies and 
quasars by using a dedicated 2.5 m telescope at Apache Point
Observatory (see Gunn et al. 2006 for technical details).
The photometric survey has imaged roughly $\pi$ steradians of the 
Northern Galactic Cap in five photometric bandpasses denoted 
by $u, g, r, i$, and $z$, and centered at 3551, 4686, 6165, 7481 and 8931 Angstroms, 
respectively. The imaging camera used is equipped with 54 CCDs 
(Fukugita et al. 1996; Gunn et al. 1998).
The limiting photometric magnitudes are
22.0, 22.2, 22.2, 21.3 and 20.5 in the previous five bandpasses,
at a signal-to-noise ratio of $5:1$. 
The median width of the PSF is $1.4''$ and 
the $\rms$ photometric uncertainties are at the $2 \%$ level
(Abazajian et al. 2004).
After image processing (Lupton et al. 2001; Stoughton et al. 2002; 
Pier et al. 2003) and calibration (Hogg et al. 2001; Smith et al. 2002),
targets are selected for spectroscopic follow-up observations.
The spectra are obtained with two dual fiber-fed CCD spectrographs at a
spectral resolution $\lambda/\Delta \lambda \simeq 1800$, and $\rms$ 
uncertainty in redshift of $\sim 30$ km$s^{-1}$.
Because of mechanical constructions, two fibers cannot be placed 
closer than $55''$ on the same tile. The incompleteness percentage of the
spectroscopic survey reaches about $6\%$. 
The SDSS spectroscopy yields three major samples: 
the Main galaxy sample (Strauss et al. 2002), the LRG sample
(Eisenstein et al. 2001), and the quasar sample (Richards et al. 2002). 
In particular, the LRG sample considered here is part of the final data release
 of the SDSS-II, indicated as DR7, which yields
$928,567$ galaxy spectra over the legacy spectroscopic coverage of $8032$ deg$^{2}$.

In this work, we made a volume-limited sample including $67,385$ LRGs 
in the redshift range from 0.16 to 0.36 and
rest-frame $\rm g$-band absolute magnitudes of $-23.2<M_{\rm g}<-21.1$,
passively evolved to $z=0.3$ (see Zehavi et al. 2005; Eisenstein et al. 2005),
by using the ``DR7-Full'' sample of Kazin et al. (2010).
K-corrections have been applied to all the galaxies in the sample, assuming a
fiducial $\Lambda$CDM model with $\Omega_{\rm m}=0.26$ and $h=1$,
not $\Omega_{\rm m}=0.25$ which was applied to the sample by 
Kazin et al. (2010).
To maximize the volume-to-surface ratio, we trim the sample 
as in Choi et al. (2010) -- see their Figure 1 for more details.
Both the Southern Galactic Cap region and the Hubble Deep Field region 
are dropped.
These cuts leave a total of $60,466$ LRGs over about
2.33 $sr$ in the survey region with an angular selection function greater
than 0.6.

Figure~\ref{fig:isosurface} shows three-dimensional 
isodensity contours of the smoothed galaxy number density fields 
obtained from the SDSS LRG sample at $\nu_f=\pm 2.0, \pm 1.5$ and 0,
of which the corresponding volume fractions are 2.3\%, 6.7\%, and 50\%.
A Gaussian smoothing is applied with $R_{\rm G}=22\hMpc$.
As expected, the asymmetry between high- and low-density regions of the 
observed genus curve shown in Figure~\ref{fig:main_result} is also clearly 
seen in this visual comparison.
The low-density regions (left upper panels) 
tend to be more connected and filamentary 
than the high-density regions (right upper panels), where structures
appear to be more isolated and rounder.
As the volume fraction increases, the structures increase in size.


\subsection{Construction of the galaxy mass and number density fields from the observed LRG sample} \label{sec:mden}

For an arbitrary large-scale galaxy survey,
the sampling of galaxies as a function of redshift is usually not uniform.
Moreover, typically the survey is designed so that
not only the mean galaxy number density is not constant, 
but also the sampling in absolute magnitude is non-uniform. 

\begin{figure}
\epsscale{1.}
\plotone{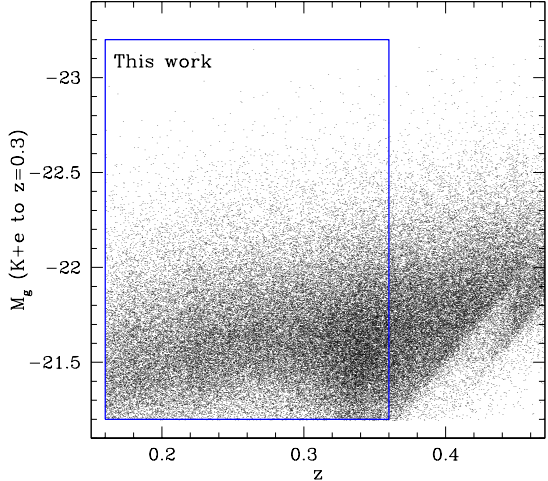}
\caption{Distribution of LRGs in the SDSS survey. 
Solid boundary lines identify the volume-limited sample used in this study,
which covers the redshift range $0.16<z<0.36$ and has rest-frame $\rm g$-band
absolute magnitudes in the interval $-23.2<M_{\rm g}<-21.1$.}
\label{fig:Mz}
\end{figure}  

An example is shown in Figure~\ref{fig:Mz}, for the semi-volume-limited 
LRG sample. The solid lines in the panel identify the volume-limited sample
considered in this study. The high non-uniformity of the sample,
as a function of redshift and absolute magnitude, is clearly visible.

In this situation, it would be incorrect to give a 
single-value weight to galaxies in each redshift bin,
based only on the radial selection function; in fact, this simple scheme would
over-weight the galaxies fully sampled, and under-weight those under-sampled. 
Galaxies with different luminosity are known to cluster differently
(Park et al. 1994; Park et al. 2005b; Zehavi et al. 2005;  Guo et al. 2013), and therefore they 
should get different weights if the sampling varies with luminosity
-- to avoid the clustering mismatch. 
The problem becomes more serious 
when galaxy luminosity or mass are used as weights, to obtain the 
galaxy luminosity density or the mass density field, respectively.
In particular, 
when the sampling in luminosity or mass varies with redshift, 
the resulting luminosity or mass density will have different mean values
across different redshift bins, even if the galaxy number density is matched.
Therefore, one should also consider the radial density gradient. 

For the construction of our galaxy mass and number density fields from
 the observed LRG sample, we devise a new weighting scheme (called `luminosity
 function matching') which properly accounts for the sampling rate variations 
 depending on the location of the galaxy both in redshift and absolute magnitude
 space, variations that are caused by the LRG target selection procedure.
In what follows, we consider the case when there is no evolution of 
the luminosity function (LF) with redshift, and briefly summarize our
procedure (see also Figs.~\ref{fig:lf}, 
\ref{fig:ndenhist} and \ref{fig:mag2mass} below).

\begin{enumerate}[i.]
\item
Select a reference redshift bin, and compute the reference LF in this bin --
indicated with $\Phi_{\rm ref}(M_{\rm g})$. 
The LF determined in this (arbitrary) redshift interval will be used to 
match the LF in other redshift bins, as shown in Figure~\ref{fig:lf}.
For our study, we choose the interval $0.16<z<0.20$ as the reference $z$-bin;
$\Phi_{\rm ref}(M_{\rm g})$ computed in this interval is indicated with the
thick solid line in Figure~\ref{fig:lf}.
\begin{figure}
\epsscale{1.}
\plotone{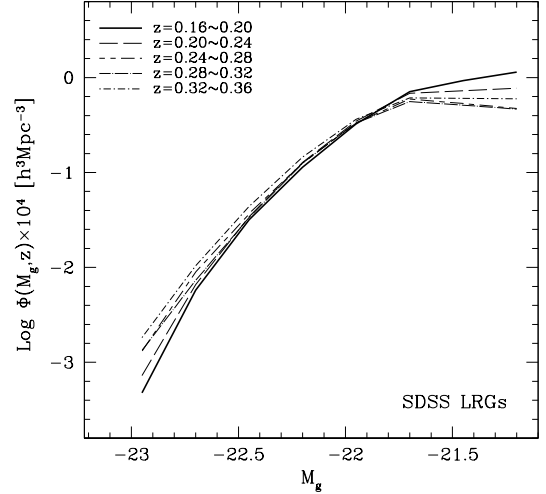}
\caption{
Luminosity function computed at different redshift bins, from the
SDSS LRG sample.  The plot is used to construct the
 galaxy mass and number density fields from the observed LRG sample, with the
  `luminosity function matching' procedure described in the main text.
 The interval $0.16 < z < 0.20$ is used as the reference redshift bin, and
  the reference luminosity function $\Phi_{\rm ref}(M_{\rm g})$ computed in this interval is indicated with the thick solid line in the figure.}
\label{fig:lf}
\end{figure}  

\item
Select bins in the two-dimensional plane defined by redshift versus
absolute magnitude, and compute the LF for each pixel of the two-dimensional
array $(M_{\rm g},z)$ above an absolute magnitude cut, $M_{\rm g,cut}$.

\item
For each pixel of the two-dimensional array $(M_{\rm g},z)$, calculate the proper
weight $w(M_{\rm g},z)$ as
\begin{equation}
w(M_{\rm g},z)=\Phi_{\rm ref}(M_{\rm g})/\Phi(M_{\rm g},z). \label{eq:w}
\end{equation}
From Figure~\ref{fig:lf}, one can easily infer, just by looking in
the magnitude range fainter than $M_{\rm g}\sim-21.9$, that this weighting scheme
will clearly weight galaxies at moderate redshifts 
more than those at low redshifts.

\item
Construct the galaxy number density field,
weighting each galaxy by $w(M_{\rm g},z)$, 
which is linearly interpolated from the two-dimensional array
$(M_{\rm g},z)$ computed as described in the previous steps.
The number density field obtained in this way 
will be uniform both in redshift and luminosity space
(see the thick blue histogram in Fig.~\ref{fig:ndenhist}),
as opposed to the one constructed by weighting each galaxy with the radial
selection function alone (dotted line in the same figure).
In particular, the density field is calculated on a mesh with cubic pixels
from a discrete particle distribution using the cloud-in-cell (CIC) 
mass assignment scheme.

\begin{figure}
\epsscale{1.}
\plotone{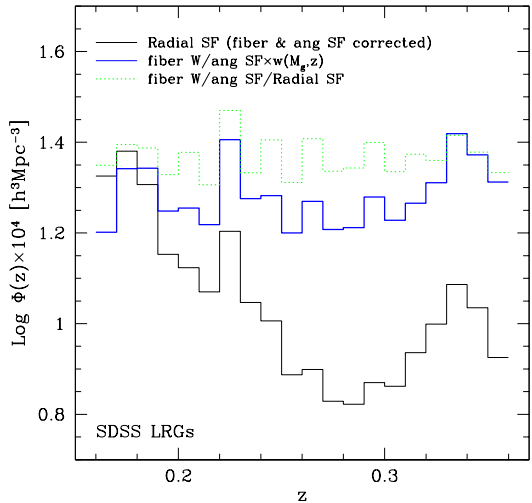}
\caption{Comoving number density of the SDSS LRGs sample as a
function of redshift (thin solid line). 
The number density has been corrected for object-loss
due to fiber collisions (Zehavi et al. 2005) and spectroscopic 
completeness. Dotted line (green histogram) 
shows the radial distribution constructed
by weighting each galaxy only with the radial selection function. Thick solid line (blue histogram)
is the galaxy number density field constructed by weighting each galaxy
with the new `luminosity function matching' weighting procedure described
in the main text -- i.e. $w(M_{\rm g},z)$, Equation~\ref{eq:w}.}
\label{fig:ndenhist}
\end{figure}  
\item

Alternatively, construct the mass-weighted halo density field 
from the observed galaxy sample. 
The galaxy mass, $M_{\rm gal}$ should be the halo mass $M_{\rm h}$ 
   corresponding to the $\rm g$-band galaxy luminosity $M_{\rm g}$, i.e.
   $M_{\rm h} = f(M_{\rm g})$ (see point iii. in Section~\ref{sec:mock} 
   for more details). Using the LRG cumulative LF measured at the reference 
   redshift bin, and the halo cumulative mass function derived from a full 
   cubic data snapshot of the HR3 
   at $z=0.2$ (which is compatible with the reference redshift), we apply the 
   halo-galaxy one-to-one correspondence model (HGC) of Kim, Park, \& Choi (2008)
   and convert galaxy luminosities into halo masses, and 
   vice-versa.


\begin{figure}
\epsscale{1.}
\plotone{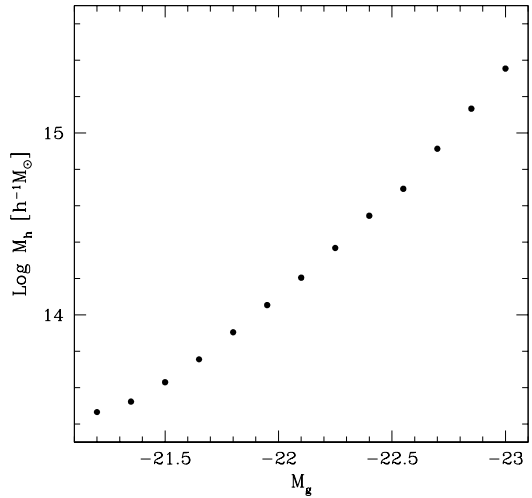}
\caption{Relation between galaxy luminosity ($M_{\rm g}$) and halo mass ($M_{\rm h}$)
obtained from the HR3 with the
halo-galaxy one-to-one monotonic correspondence model (HGC)
of Kim, Park, \& Choi (2008). The mapping  is used
to compute the galaxy mass $M_{\rm gal}$  (see the end of Sec.~\ref{sec:mden}).}
\label{fig:mag2mass}
\end{figure}


Figure~\ref{fig:mag2mass} shows the relation between galaxy luminosity and 
halo mass, used to determine $M_{\rm gal}$.
The halo mass corresponding 
to the absolute magnitude cut, $M_{\rm g,cut}$, is 
$M_{\rm h,cut}=10^{13.466} h^{-1} M_\odot$.
To compute the galaxy mass density field, each galaxy is weighted by 
$w(M_{\rm g},z)\times M_{\rm h}$.
The mass density field derived with this procedure is
equivalent to the one obtained from uniformly selected LRGs.
\end{enumerate}



\section{Simulated LRG Samples: Description and Methodology}  
\label{sec:sims_methods}

\begin{figure*}
\epsscale{1.2}
\plotone{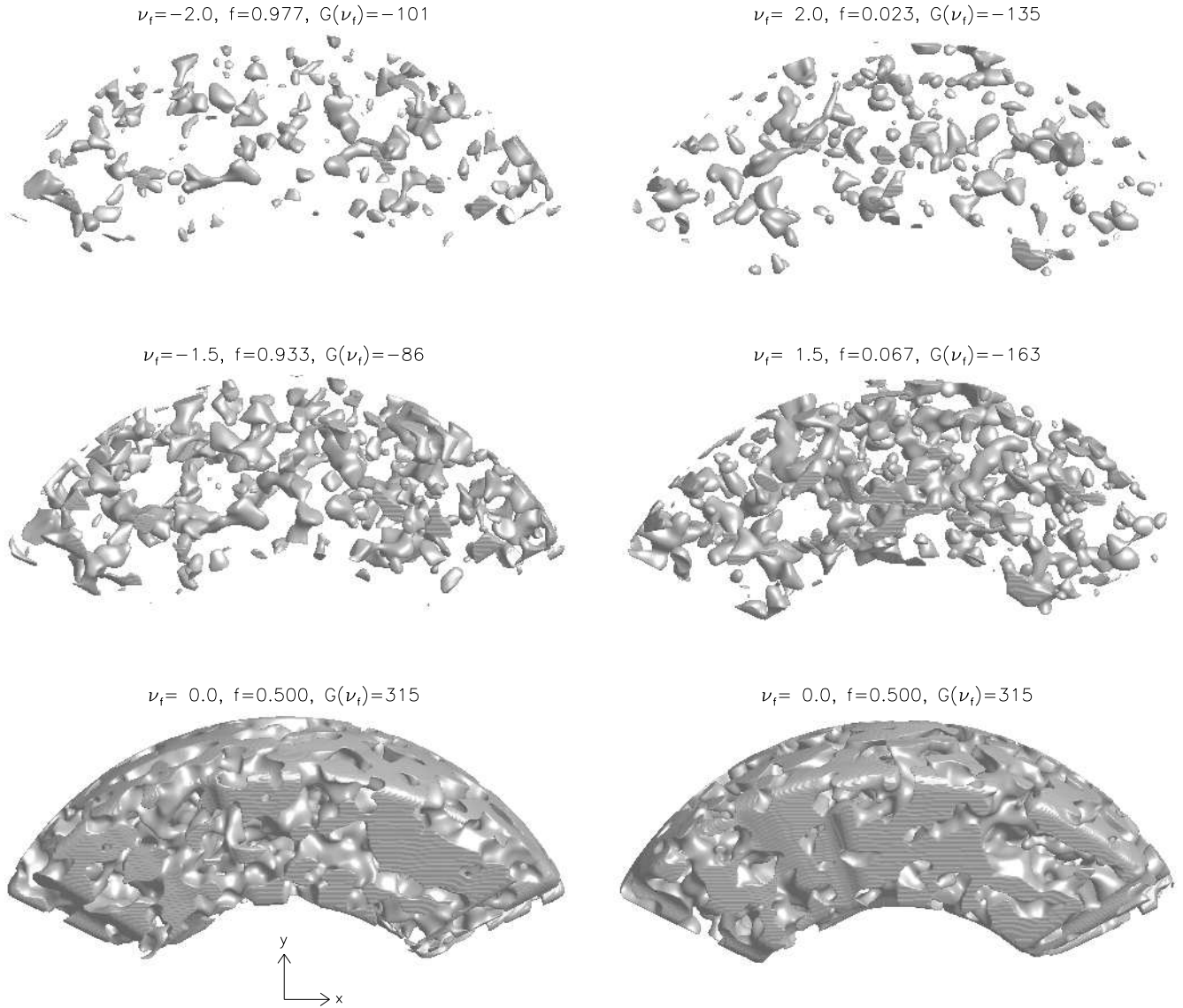}
\caption{
Three-dimensional view of the galaxy number density field of a mock
sample smoothed with $R_{\rm G} = 22\hMpc$.}
\label{fig:isosurface_mock}
\end{figure*}


In this section we briefly describe the Horizon Run 3 $N$-body simulation, 
and the procedure to construct the SDSS DR7 mock LRG samples from the 
simulation output. We will then compare numerical results and  measurements 
from data in Section \ref{sec:genus_top}.
The mock surveys  will also be used to quantify several nonlinearities
due to systematics which affect the genus curve: correcting for these effects 
allows one to accurately recover the topology of the underlying matter, as we 
will present in Section \ref{sec:genus_sys}.

\subsection{The Horizon Run 3 $N$-body simulation}

The Horizon Run 3 (HR3; Kim et al. 2011)
is one of the largest $N$-body simulations to date,
made using $7210^3 = 374$ billion particles, spanning a volume
of $(10.815~h^{-1}$Gpc)$^3$ -- which is
over 8800 times the volume of the Millennium Run (Springel et al. 2005).
The particle mass is down to $2.44 \times 10^{11}h^{-1}M_{\odot}$,
allowing to resolve galaxy-size halos with mean particle separation of $1.5~h^{-1}$Mpc.
The simulation is based on the $\Lambda$CDM cosmology, with parameters fixed by the 
WMAP 5-year data (Komatsu et al. 2009). The
linear power spectrum used is obtained with the CAMB source
(http://camb.info/sources), which provides a better measurement of the BAO scale.
The simulation starts at $z_{\rm i} = 27$, and is evolved till the present epoch
with 600 global time steps. The code used for the run is an improved version of the 
Grid-of-Oct-Trees-Particle-Mesh code (GOTPM), originally devised by
Dubinski et al. (2004);  a new procedure has been implemented, in order to
describe more accurately the particle positions using single precision.

In the HR3, halos are first identified via a standard Friend-of-Friend (FoF) procedure. Then subhalos are found  -- out of FoF halos -- with a subhalo
finding technique developed by Kim \& Park (2006) and Kim, Park, \& Choi (2008).
This method allows one to identify physically self-bound (PSB) dark matter subhalos not tidally disrupted by larger structures at the desired epoch.
In particular, LRGs are identified as the most massive dark matter subhalos.
To make the comparison with observational data,
we saved the particle positions and velocities along the past
light cone for 27 separated observers, and found  subhalos
in the past light cone surface from $z=0$ to $z=0.7$.

From each simulated light cone, we made 3 mock samples using exactly
the same survey mask and angular selection function of the SDSS sample.
In addition, we
applied the same smoothing length as for the observational case.
In total, we are able to obtain
81 non-overlapping mock samples, thanks to the
enormous volume of the HR3.
To this end, we note that the ability to
simulate big volumes is essential (particularly for the LRG distribution), since larger volumes allow one to model more accurately the true power at large scales and the corresponding power spectrum.
A large box size will guarantee small statistical errors in power spectrum estimates, so that the acoustic peak scale can be
measured with an accuracy better than $1\%$, and the genus curve characterized with
unprecedented  statistical significance.


\subsection{Construction of the mock LRG samples}\label{sec:mock}


A crucial step in our analysis is the construction of  realistic mock
LRG samples.
This requires the ability to mimic all the observational biases, such as
survey boundary, radial and angular selection function, redshift space 
distortions and so forth. To build the various simulated  catalogs,
27 observers were placed in the HR3 box, each covering the redshift 
range $0 < z < 0.7$ without overlaps: this means that the survey volumes are 
totally independent.
In addition, the LRG mocks are made so that they span exactly the same range 
in absolute magnitude as the observational sample;
hence, the number of galaxies in each mock is nearly equal to the observed one
(at the percentage level accuracy). Moreover,
the simulated galaxies should be observed in redshift space
along the past light cone with the same radial and angular selection function 
of the observational sample,
and also with the same selection function in absolute magnitude space
as in the real observation.

Figure \ref{fig:isosurface_mock} shows a
three-dimensional example of the simulated LRG number density field 
obtained from the HR3, smoothed with a Gaussian
filter at $R_{\rm G} = 22\hMpc$ scale. The plot
is the equivalent of Figure \ref{fig:isosurface}, but now
for the LRG mock samples constructed from the HR3 simulation.
Again, the left panels display
three representative density contours enclosing low-density
regions, which occupy respectively $ 2.3\%$ ($\nu_{\rm f}=-2.0$; top), 
$6.7\%$ ($\nu_{\rm f}=-1.5$; intermediate),
and $50.0\%$ ($\nu_{\rm f}=0.0$; bottom) of the sample volume.
The same thresholds, but now with positive signs and so for
high-density
regions  (i.e. $\nu_{\rm f}=2.0$, top; $\nu_{\rm f}=1.5$, 
intermediate; $\nu_{\rm f}=0.0$, bottom), are shown in the right panels.

\begin{figure}
\epsscale{1.}
\plotone{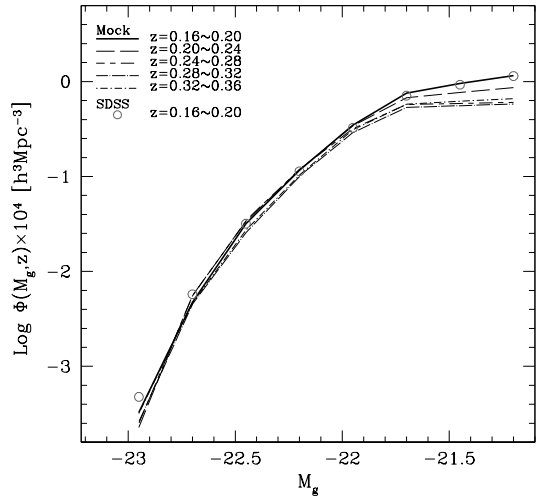}
\caption{Luminosity function computed at different redshift bins from one of our LRG mocks, as indicated in the panel.
Open circles in the figure are
luminosity function measurements derived from the SDSS LRG sample in the reference redshift bin $0.16 < z < 0.20$ -- as in Figure \ref{fig:lf}.
Clearly, simulated results agree very well with observational
measurements.}
\label{fig:lfmock}
\end{figure}

\begin{figure}
\epsscale{1.}
\plotone{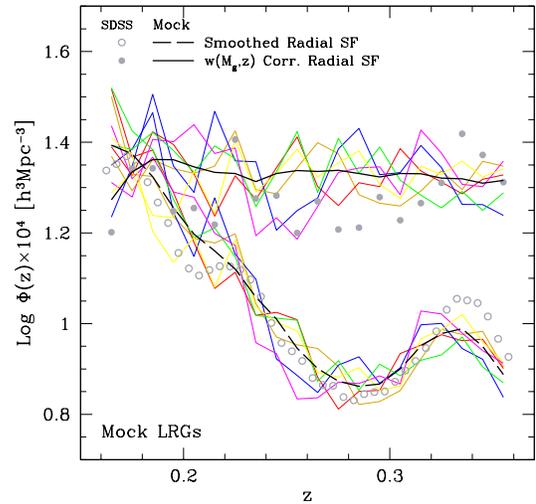}\caption{
Comoving number density of mock LRGs averaged over all the mock samples as a
function of redshift (black dashed line). Solid line shows the radial distribution
constructed by weighting each galaxy by $w(M_{\rm g},z)$. 
Colored thin lines show
results from six mock surveys. For comparison, the observational results are also
plotted (circles).}
\label{fig:ndenhist_mock}\end{figure}

Since we apply identical techniques both to the
SDSS LRG sample and to the LRG mock surveys, we expect the results of the 
analysis to be identical across datasets -- within statistical variations --
if the simulations are correctly modeling the distribution of galaxies.
In what follows, we describe in more detail how to
build the SDSS DR7 mock LRG samples from the HR3 simulation output.
Results from our procedure
confirm that we are correctly modeling the LRG distribution
(see Figs.~\ref{fig:lfmock} and \ref{fig:ndenhist_mock}).
The major steps of the construction process are summarized next.


\begin{enumerate}[i.]
\item
Locate 27 observers in the HR3 simulation box, and save all dark halos along 
the past light cone of the observer during the simulation, in the redshift
range $0<z<0.7$. 
From each light cone data, make 3 mock surveys using exactly
the same survey mask and angular selection function as for the SDSS volume-limited
sample.

\item
Apply a proper correction to make the halo mass function uniform in redshift.
In fact, in the HR3 simulation the minimum mass limit of subhalos
       that can have LRGs with constant observed number density varies as 
       a function of redshift (see Fig.~6 in Kim et al. 2011).
This leads to the following
relation; $f(z)=(-8.743\times 10^{12}h^{-1}M_{\odot})z+
(1.711\times10^{13}h^{-1}M_{\odot})$. 
The correction one needs to apply to the halo mass at an arbitrary redshift
$z$ is then given by the ratio $f(z_{\rm ref})/f(z)$, where 
$f(z=z_{\rm ref})=1.55\times 10^{13}h^{-1}M_{\odot}$ is the minimum halo mass at
a median redshift of $z_{\rm ref}=0.18$ in the reference redshift bin
(recall the procedure described in Section~\ref{sec:mden}, 
and the chosen reference redshift interval).

\item
Populate dark matter halos with galaxies using a suitable correspondence
scheme. In essence, to connect galaxies with halos
one needs to make an assumption on the relation between galaxy luminosity
and halo mass.
A widely-used approach is the subhalo abundance matching,
where more luminous galaxies are assigned to more massive haloes
(Kravtsov et al. 2004; Tasitsiomi et al. 2004; Vale \& Ostriker 2006;
Conroy \& Wechsler 2009; Guo et al. 2010; Behroozi et al. 2010; 
Kim, Park \& Choi 2008).
This scheme assumes that halos with mass above a certain threshold and
with a given mean number density correspond to galaxies with
luminosity or mass above a certain threshold and having the same mean halo number
density. 
For our mocks, we apply the  halo-galaxy one-to-one monotonic correspondence 
model (HGC)
of Kim, Park, \& Choi (2008) which extends the subhalo abundance matching procedure:
there is one and only one galaxy in each subhalo, and a more massive subhalo hosts a more luminous galaxy.
 The mapping $M_{\rm h} = f(M_{\rm g})$ is shown in Figure \ref{fig:mag2mass}.
This correspondence scheme allows us to
assign a luminosity to each LRG mock galaxy, and to compute galaxy masses 
(see also the end of Sec.~\ref{sec:mden}).

\item
Account for the effects of the color-dependent
luminosity cut imposed by the SDSS LRG volume-limited sample
selection criteria, both in redshift and luminosity space, 
which reduces the sampling density (see again Sec.~\ref{sec:mden}).
In order to do so,
we discard mock galaxies with a rejection probability given by
$1/w$, where $w \equiv w(M_{\rm g}, z)$ is derived from the observed sample 
(see Eq.~\ref{eq:w}).

\end{enumerate}

Our procedure successfully reproduces the dependence of the LRG sampling rate
on luminosity and redshift as in the SDSS LRG sample. This is
shown in Figures~\ref{fig:lfmock} and \ref{fig:ndenhist_mock}, 
the corresponding counterparts of
Figures~\ref{fig:lf} and \ref{fig:ndenhist}, respectively, 
obtained from simulated samples.
In particular,  Figure~\ref{fig:lfmock} displays the luminosity function 
computed at different redshift bins from one of our LRG mocks. 
To facilitate the comparison with the actual SDSS data, 
open circles in the figure are luminosity function measurements 
derived from the SDSS LRG sample in the reference redshift bin 
$0.16 < z < 0.20$ -- as in Figure~\ref{fig:lf}, see Section~\ref{sec:mden}.
Clearly, the simulated results agree very well with observational 
measurements in terms of matching the luminosity function.
To this end, Figure~\ref{fig:ndenhist_mock} shows the comoving number density 
of SDSS mock LRGs averaged over all the 81 mock samples,
as a function of redshift (black dashed line). Solid line shows the
radial distribution obtained by weighting each galaxy with $w(M_{\rm g},z)$.
Colored thin lines show results of six arbitrary mock surveys.
Filled and open circles in the figure are
analogous measurements derived from the SDSS LRG sample
-- as in Figure \ref{fig:ndenhist}.
The weighting scheme used is explained in Section \ref{sec:mden}.
Even in this case, 
the plot confirms the correctness of our modeling procedure: our mocks 
have the same sampling rate in redshift as the observed
SDSS LRG sample.



\section{Genus Topology of LRGs: SDSS versus Mock Measurements} 
\label{sec:genus_top}

In this section we present results for the genus measured from the SDSS DR7 
LRG sample, and from our LRG mocks obtained from the HR3 LCDM  simulation. 
In both cases, we compute the genus curves using the
mass weighted density field and the number density field -- although later on 
we will only use the number density.
By contrasting observational results against mock measurements which assume 
Gaussian initial conditions,
we detect significant non-Gaussian deviations of the observed genus curve 
from theoretical expectations.
We then further quantify these discrepancies by introducing a new statistical test.
A large part of the non-Gaussian deviations is caused by systematics, and
we address their impact on the genus curve in Section \ref{sec:genus_sys}. 

\subsection{Genus of SDSS LRGs from the mass weighted density field and from the number density field} \label{sec:weight}

\begin{figure}
\epsscale{1}
\plotone{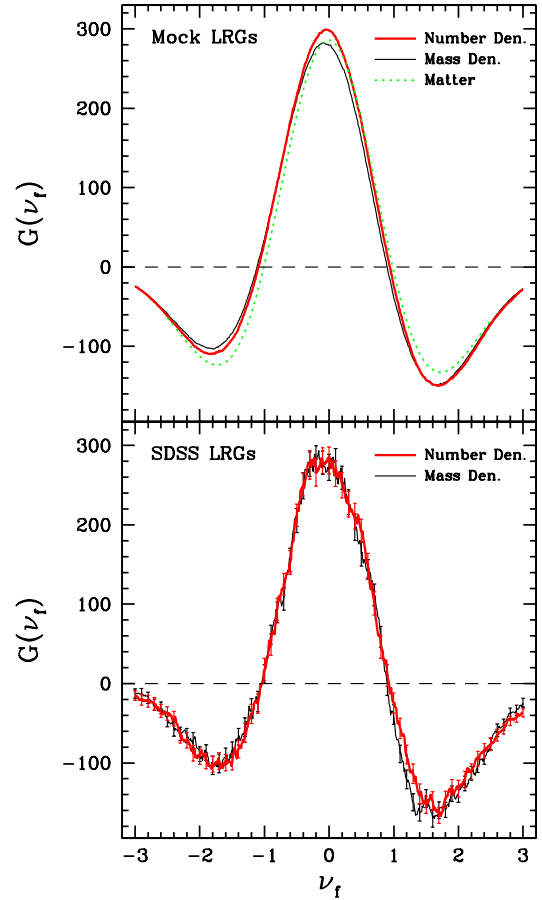}
\caption{
Genus curves derived from the smoothed mass density and number density fields, 
respectively, as a function of the volume threshold $\nu_{\rm f}$.
Measurements are not corrected for systematics yet.
A Gaussian smoothing of
radius  $R_{\rm G}=22\hmpc$ has been applied, at a pixel size of $p=3.7\hmpc$.
In both panels, red thick solid lines are used for the number density field, and 
black thin lines for the mass weighted density field.
(Top) Measurements
averaged from 81 mock LRG samples. The genus curve of the dark matter
distribution in real space using the full simulation cube is also shown,
with the dotted green line.
(Bottom) Analogous
measurements from the SDSS sample. See the main text for more details.}
\label{fig:mgplot}
\end{figure}

In Sections~\ref{sec:mden} and ~\ref{sec:mock} we constructed the
mass weighted density field and the number density field in order to compute the genus.
This is because
Jee et al. (2012) found that the halo mass density has 
a much tighter (and simpler) relation with the underlying matter density
than the halo number density. A similar conclusion was reached by 
Park, Kim, \& Park (2010), in relation to the gravitational shear field.
To this end, Figure~\ref{fig:mgplot} shows the genus curves derived from the
mass density and number density fields, as a function of the volume fraction,
$\nu_{\rm f}$.


No corrections for systematics are yet applied.
We smooth our density field (either the mass weighted or the number density one)
with a Gaussian filter of radius
$R_{\rm G}=22\hmpc$ at a pixel size of $p=3.7\hMpc$.
To minimize any nonlinearity introduced by the choice of the pixel dimension, 
we use the smallest possible pixel size we can afford.
In the figure, red thick solid lines are used for the number density field, 
and black thin lines are for the mass weighted density field.

In particular, the top panel shows our measurements
averaged from 81 mock LRG samples, 
where we also plot the genus curve of the  dark matter
distribution in real space using the full simulation cube  --
with the dotted green line. 
The difference between halo and matter density field curves are mostly due to
halo biasing and discrete sampling of the halo density field (i.e. shot noise).
The genus amplitude measured from the halo mass weighted density field
     is lower than the one obtained from the halo number density field.
This is consistent with  
the result of Seljak, Hamaus, \& Desjacques (2009); namely,
weighting halo galaxies by halo mass can significantly 
suppress shot noise. 

The bottom panel shows similar measurements from the SDSS LRG sample.
To date, these are the
best measurements of the genus curve from the SDSS survey catalog.
In Table~\ref{tab:mgplot_genus} of Appendix~\ref{sec:genus_curves}, the genus values are
given as a function of the volume-fraction threshold level.
In particular, the genus amplitude obtained from the number density field
is equal to $G_{\rm o}=285$, with a $\sim 4.0\%$ error
including all systematic effects such as finite pixel size, survey boundary,
radial and angular selection functions and sparse sampling 
(see also Fig.~\ref{fig:main_result}).
The errorbars in the figure are the $1\sigma$ deviations
computed from 81 independent mock samples.


\subsection{Genus-related statistics: quantifying the non-Gaussian deviations} 
\label{sec:ngd}

Although the shape of the  genus curve does not
depend on the weighting scheme as much as its amplitude,
results from simulations reveal that the mass weighted density field tends 
to show more `meatball'-shifted topology
and more isolated clusters.
In Table \ref{tab:mgplot} we report the details of the  genus-related statistics 
(see Sec.~\ref{sec:grs}), for the genus measurements displayed in 
Figure \ref{fig:mgplot}, while 
in Appendix \ref{sec:genus_curves}
we list the complete genus values relative to these measurements, as a function of the volume-fraction threshold $\nu_{\rm f}$ (see Tab.~\ref{tab:mgplot_genus}).
We also provide similar measurements as a function of $\nu$.
The latter table may be useful for readers who wish to directly 
use galaxy clustering topology for cosmological applications.


\begin{table}
\caption{Genus-related Statistics for the LRG observed and
simulated samples considered in Figure~\ref{fig:mgplot}.}
\begin{center}
\begin{tabular}{lllll}
\hline
\hline
Sample     &  $G_{\rm fit}(0)$ & $\Delta\nu_{\rm f}$  & $A_{\rm V}$ & $A_{\rm C}$ 
\\ \hline
DM    & $285.1$ &$ -0.009$ &$ 0.97$ &$1.05$ \\
\hline
\multicolumn{2}{l}{Mock LRGs}\\
Number& $299.0\pm11.5$ &$ -0.058\pm 0.018$ &$ 0.79\pm0.06$ &$1.14\pm0.06$ \\
Mass & $280.7\pm11.0$ &$ -0.086\pm0.019$ &$ 0.77\pm0.05$ &$1.21\pm0.05$ \\
\hline
\multicolumn{2}{l}{SDSS LRGs}\\
Number & $285.2$ & $-0.047$ &$0.79$& $1.22$\\
Mass & $284.5$ &$-0.056$ & $0.74$& $1.33$\\
\hline
\end{tabular}
\end{center}
{\bf Notes.} 
`Number' and `Mass' stand for number density field and mass weighted density field,
respectively.
$G_{\rm fit}(0)$ is the amplitude of the best-fit Gaussian genus curve, $\Delta \nu$ is the
shift parameter, and $A_{\rm C}$ and $A_{\rm V}$ are cluster and void abundance parameters,
respectively. All these values are not bias-corrected.
Uncertainty limits are estimated for 81 mock samples. 
Deviations of the genus curves from the Gaussian expectation are
quantified by $\Delta\nu_{\rm f}$, $A_{\rm V}$ and $A_{\rm C}$.
\label{tab:mgplot}
\end{table}

\begin{figure}
\epsscale{0.65}
\centering
\plotone{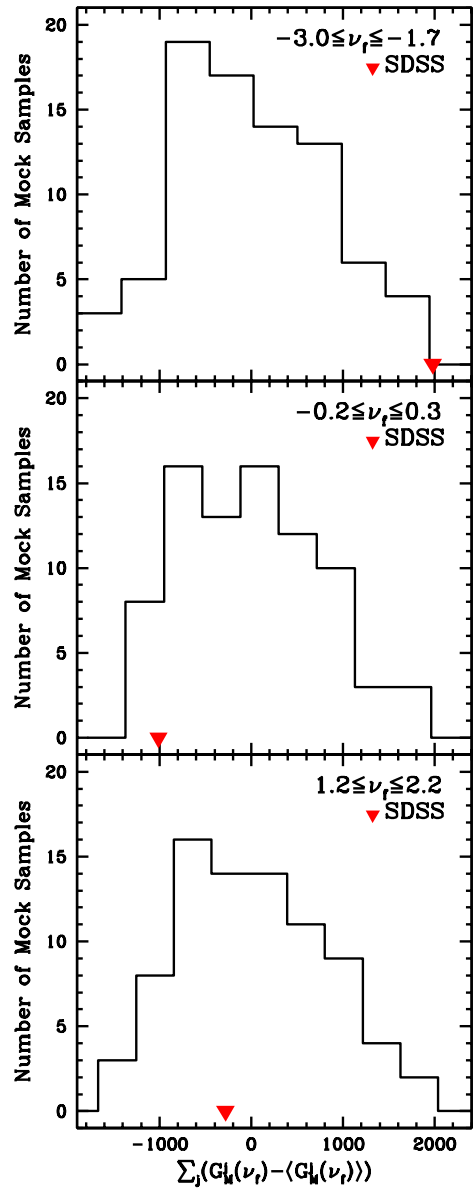}
\caption{
Statistical test to estimate
quantitatively the non-Gaussian discrepancies between the predicted and observed genus curve, for the SDSS sample -- as explained in the main text.
Histograms are integrals of the
the differences between the genus curve of each individual mock sample
and the curve obtained by averaging  all our 81 mock LRG samples.
Three different intervals in $\nu_{\rm f}$ are considered, as indicated in the various panels.
Triangular symbols are measurements obtained from the SDSS sample
at the corresponding threshold
intervals. 
The discrepancies are seen near the mean density regions and  
       in low-density regions ($-3.0\le\nu_{\rm f}\le-1.7$), with a significance 
       level of 90\%.}
\label{fig:gstats_nden}
\end{figure}

From Figure~\ref{fig:mgplot} and from the results for the genus-related 
statistics, it is evident that in the mock measurements
low-density regions are definitely less affected by the weighting scheme.
Instead, the weighting scheme does not introduce any change in the genus
amplitude derived from the observational sample, and the mass
weighted density field still produces more isolated clusters.
This finding clearly points towards the existence of a significant amount of
scatter in the relation between galaxy observables and their underlying halos;
hence, one needs to gain a better understanding of the connection between
galaxies and their dark matter halos, and of the galaxy formation process 
in general. In this paper, hereafter 
we will adopt the use of number density instead of the mass weighted density,
for simplicity.

Overall, the HGC galaxy assignment scheme of Kim, Park, \& Choi (2008) 
    is able to match well the observed amplitudes and shapes of the corresponding genus curves.
In fact, from Table~\ref{tab:mgplot}, one can notice that
the values obtained for $\Delta \nu_{\rm f}$, $A_{\rm V}$ and $A_{\rm C}$
from the simulated LRG samples agree well -- within the quoted uncertainties 
-- with those measured from the observational sample.
However, we detect some significant discrepancies between mocks and observations 
for $A_{\rm V}$ in the lower density regions beyond the
integration intervals quoted in Section~\ref{sec:grs} (i.e. $\nu_{\rm f}\leq -1$)
-- see indeed the difference between the red and grey lines in 
Figure~\ref{fig:main_result}.
We also detect some discrepancies for the observed and predicted genus amplitudes.

To estimate quantitatively the
statistical significance of these discrepancies, we use the following method.
First, we calculate the differences between the genus curve of each individual 
mock sample and the curve obtained by averaging all the 81 mock samples; 
we do this at three different intervals, i.e.
$-3.0\le\nu_f\le-1.7$, $-0.2\le\nu_f\le0.3$, and $1.2\le\nu_f\le2.2$.
We then plot the integrals of these differences as histograms in 
Figure~\ref{fig:gstats_nden}.
Finally, we place in the same plot our measurements obtained
from the SDSS sample at the corresponding threshold
intervals, indicated with triangular symbols. 
Here we measured genus curves from the number density fields 
of the samples.

As one can infer from the figure (with a significance level of 90\%),
        departures from Gaussianity are seen near the mean density regions, and 
        in low-density regions (i.e. in the interval $-3.0\le\nu_{\rm f}\le-1.7$).
The radical difference between the genus curves of the observational and 
simulated data in low-density regions shows that topology is 
highly sensitive to the connectivity of voids.
In the next part, we will address the key role played by systematics 
on the genus curve which will explain some of the discrepancies,
and show how to accurately correct for their effects to recover the
topology of the underlying matter.
In a forthcoming paper, we provide
an interpretation of  the remaining deviations (i.e. after correcting for known systematics)
in the context of primordial non-Gaussianity.




\section{Genus Topology of LRGs: Systematics} \label{sec:genus_sys}

In this section we briefly discuss
the known systematics which affect the genus curve.
We then test and quantify their impact on the genus using the genus-related statistics presented in Section \ref{sec:grs},
with the help of mock LRG samples  (Section \ref{sec:mock}),
and show how to correct for their effects.
By applying those corrections, we
obtain the most accurate constraint on the genus amplitude to date,
which significantly improves on our previous measurements.
In particular, Figures \ref{fig:gplot} and \ref{fig:gplotdiff}
are among the most important results of our paper.


\subsection{Impact of systematics on the genus curve}

As anticipated in Section \ref{sec:grs},
even if the initial conditions were perfectly Gaussian,
small deviations from Gaussianity are expected because of systematics.
Since systematics directly impact the shape and amplitude of the genus curve, it is imperative
to be able to quantify and correct for their effects.
This can now be done quite accurately, with the help of realistic mock catalogs such as those constructed from the HR3
(Section \ref{sec:mock}).

Broadly speaking, systematics that cause non-Gaussian deviations in the genus curve can be classified into three main classes:
those due to the observational or analysis strategy, those due to statistics, and those of cosmological origin.
Finite pixel size effects, survey boundary mask, radial and angular selection
function, past light cone gradient, and initial conditions of the simulations belong to the first class.
Shot noise or sparse sampling and cosmic variance are of statistical origin, while
galaxy biasing, nonlinear gravitational evolution, and redshift-space distortions (RSD) are related to cosmology.
Sometimes, but this depends on the chosen terminology,  the last class is not considered as a systematic effect.
Here, we broadly term all these three classes as systematic biases on the genus curve.

Clearly, several of the previously mentioned effects are connected, and so
one needs to remove them simultaneously.
In the absence of other known systematics, an eventual residual of non-Gaussianity (after applying the corrections mentioned above)
has to be ascribed to a primordial origin.
In what follows, we discuss in particular the nonlinear gravitational evolution, and the effects of galaxy bias and past light cone on the genus.
More details on the full modeling of systematics in topology measurements
      will be presented in Young-Rae Kim et al. (in preparation).


\subsection{Modeling and correcting for systematics} 

\begin{figure}
\epsscale{1.0}
\epsscale{1.0}
\plotone{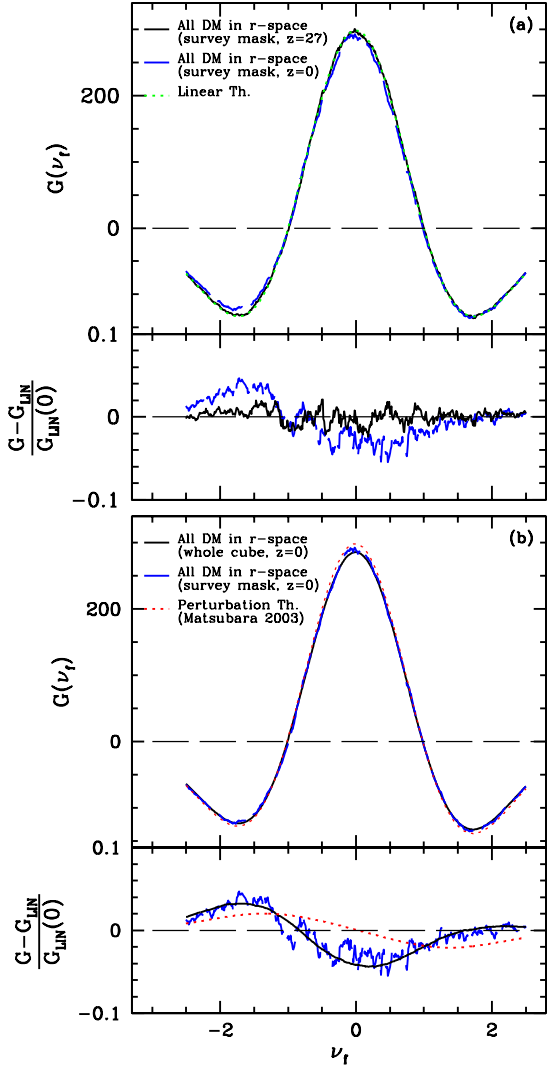}
\caption{
Impact of systematics on the genus curve: cosmic variance, SDSS survey mask,
initial conditions of the simulations, and gravitational evolution.
A Gaussian smoothing length of $R_{\rm G}=22\hMpc$ is applied.
(a) The black solid line shows the genus curve averaged over all the 81
initial matter density fields at $z_{\rm i}=27$, in real space.
The blue dashed line is the corresponding final one, at $z_{\rm f}=0$, 
computed similarly.
The SDSS survey mask is applied.
The green dotted curve is the predicted linear theory genus, 
relative to the entire survey volume.
The lower part in the same panel displays the ratio
$\Delta = [G(\nu_{\rm f})-G_{\rm LIN}(\nu_{\rm f})]/G_{\rm LIN}(\nu_{\rm f}=0)$,
as a function of the volume threshold $\nu_{\rm f}$ and for the two different 
redshifts considered. See the main text for more details.
(b) Genus curve of the matter distribution in real space at $z=0$, 
using the full cubic data.
The black solid line is used for the full simulation cube, 
while the dashed blue line is obtained with the SDSS LRG survey mask applied.
The dotted red line is the second-order perturbation theory prediction
of Matsubara (2003).
The difference between
the red and black lines shows the discrepancy between
second-order perturbation theory prediction and $N$-body simulations.}
\label{fig:sysbias_dm}
\end{figure}

\begin{figure}
\epsscale{1.0}
\plotone{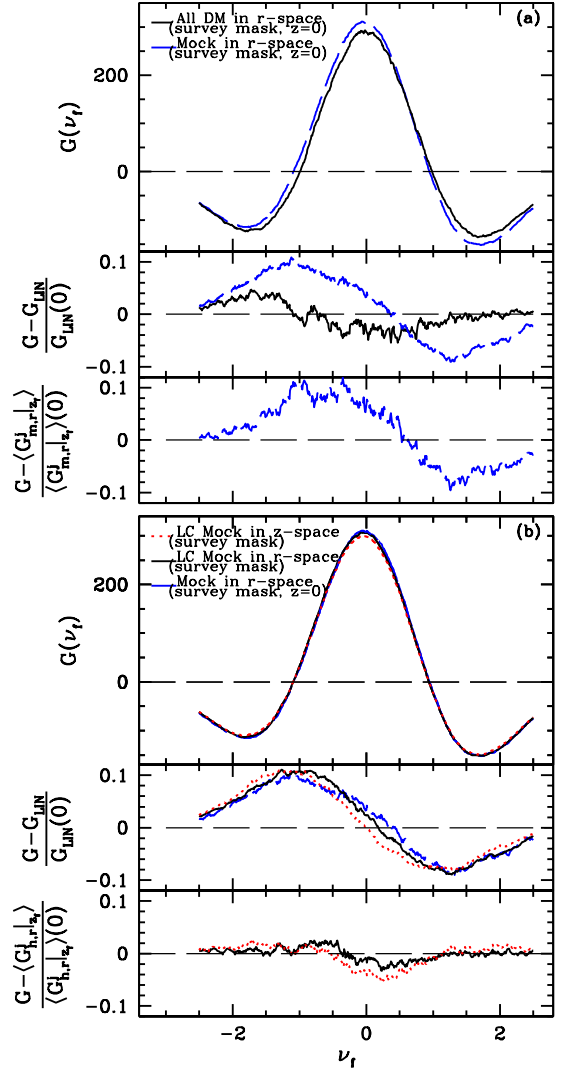}
\caption{
Impact of systematics on the genus curve: galaxy (halo) biasing, shot noise, 
past light cone gradient and RSD. A Gaussian smoothing 
length of $R_{\rm G}=22\hMpc$ is applied.
(a) Effects of shot noise and galaxy biasing.
The black solid line shows the genus obtained
from the averaged matter density field, and the dashed blue line 
is used for the one obtained from an average of the halo density field.
Similarly as in Figure~\ref{fig:sysbias_dm}, the middle part in the same panel
displays the quantity $\Delta$, which proves that
the dark matter density field and the halo number
density have very different topology.
In the bottom part of the same panel, 
$\langle G^{\rm j}_{\rm m,r}|z_{\rm f}\rangle$ is the genus value averaged 
over all the 81 mocks in real space 
at the final epoch.
(b) Effects of
past light cone gradient and RSDs.
The solid black and dotted red lines are the genus curves averaged over 81 
past light cone mock SDSS surveys of the LRGs in real and redshift spaces, 
respectively. See the main text for more details.
In the lower part of the same panel,
${\langle}G^{\rm j}_{\rm h,r}|z_{\rm f}{\rangle}$ is the genus value averaged
over all the 81 halo density fields in real space at the final epoch.}
\label{fig:sysbias_halo}
\end{figure}


In this work we perform similar corrections as those applied by Choi et al. (2010), who studied the effects of systematics on the genus computed from the
nearby Main galaxy sample of the SDSS DR7.
The overall goal is to remove the nonlinear systematics in the observed sample step-by-step, as well as
to estimate the genus curve of the underlying matter density field
using a set of mock samples.

We first consider the effect of nonlinear gravitational evolution on the 
genus curve, measured from our mock LRG samples -- 
along with survey mask and initial condition effects.
For this purpose, we compute the genus curves of 81 dark matter density 
fields, both at the initial ($z_{\rm i}=27$) and final ($z_{\rm f}=0$) redshifts; 
these quantities, measured in real space,
are indicated as $G^{\rm j}_{\rm m,r}|{z_{\rm i}}$ and 
$G^{\rm j}_{\rm m,r}|{z_{\rm f}}$, respectively, 
where the index $j$ refers to the particular light-cone mock survey 
considered, $m$ stands for matter, and $r$ for real space.
Each density field is selected within the SDSS survey mask, at
a particular region in the simulation so that the evolved density field 
has a one-to-one correspondence with the initial density field.


\begin{table}
\caption{Genus-related Statistics of the Samples used in
Figures~\ref{fig:sysbias_dm} and \ref{fig:sysbias_halo}.
}
\begin{center}
\begin{tabular}{llrrr}
\hline
Genus                    &  $G_{\rm fit}$ & $\Delta\nu_{\rm f}$  & $A_{\rm V}$ & $A_{\rm C}$ \\ \hline
$~G_{\rm m,r}|z_{\rm f}$\footnote{Genus value of the $z=0$ snapshot matter distribution in
real space using the full cubic data.}         & $285.1$ &$ -0.009$ &$ 0.97$ &$1.05$ \\
${\langle}G^{\rm j}_{\rm m,r}|z_{\rm i}{\rangle}$\footnote{Genus value averaged
over all the 81 dark matter density fields in real space at the initial epoch $z=27$.}
& $288.0$ &$ -0.007$ &$ 0.95$ &$1.05$ \\
${\langle}G^{\rm j}_{\rm m,r}|z_{\rm f}{\rangle}$\footnote{Genus value averaged
over all the 81 dark matter density fields in real space at the final epoch $z=0$.}
& $297.1$ &$  0.003$ &$ 0.99$ &$1.10$ \\
${\langle}G^{\rm j}_{\rm h,r}|z_{\rm f}{\rangle}$\footnote{Genus value averaged
over all the 81 halo density fields in real space at the final epoch.}
& $310.1$ &$ -0.044$ &$ 0.80$ &$1.12$ \\
${\langle}G^{\rm j}_{\rm h,r}|{\rm LC}{\rangle}$\footnote{Genus value averaged
over all the 81 past light cone mock galaxy density fields in real space.}
& $307.0$ &$ -0.057$ &$ 0.79$ &$1.13$ \\
${\langle}G^{\rm j}_{\rm h,z}|{\rm LC}{\rangle}$\footnote{Genus value averaged over
all the 81 past light cone mock galaxy density fields in redshift space.}
& $299.0$ &$ -0.058$ &$ 0.79$ &$1.14$ \\
\hline
\label{tab:sysbias}
\end{tabular}
\end{center}
\end{table}


Panel (a) in Figure \ref{fig:sysbias_dm} shows the gravitational evolution 
effect on the genus curve, in real space.
The black solid line is the genus curve averaged over all the 81 simulated initial 
matter density fields (at $z_{\rm i}=27$),
while the blue dashed line is the corresponding final one, at $z_{\rm f}=0$, 
computed in the same way.
The green dotted line is the predicted linear theory genus, relative to the entire 
survey volume.
The SDSS survey mask is applied. 
The lower part in the same panel displays the ratio
$\Delta = [G(\nu_{\rm f})-G_{\rm LIN}(\nu_{\rm f})]/G_{\rm LIN}(\nu_{\rm f}=0)$,
as a function of the volume threshold $\nu_{\rm f}$ and for the two different 
redshifts considered.
In particular, the shape of $\Delta(\nu_{\rm f})$ at $z_{\rm i}=27$   
      is affected by cosmic variance -- which causes small deviations from Gaussianity (of statistical nature)
in a finite volume sample -- and bias, which arises from the initial conditions of the HR 
      simulations obtained via the Zeldovich approximation.
The genus curves averaged over all the
matter density fields are relatively noisy (about $9\%$).

Panel (b) in Figure \ref{fig:sysbias_dm} shows instead
the genus curve of the matter distribution in real space at $z=0$, 
using the full cubic data.
The black solid line is used for the full simulation cube, 
while the dashed blue line is obtained with the SDSS LRG survey mask applied.
The ratio $\Delta{\nu_{\rm f}}$ gives information only 
  about nonlinear gravitational evolution.
We have attempted to fit the discrepancy with the 
  second-order perturbation theory prediction of Matsubara (2003). 
  The dotted red line in the lower panel of the same figure shows the result of this fit.
  Line colors and styles are the same as in the upper part of the panel.
The skewness parameters from the nonlinear gravitational evolution,
$S_{\rm gr}^{(\rm a)}$, are calculated by integrating the bispectrum
of the matter density distribution $B_{\rm gr}$
given in terms of the second-order
correction to the density fluctuations from nonlinear gravitational
clustering in the weakly non-Gaussian regime (see the equations in Section 4.2 
of Matsubara 2003):
$S_{\rm gr}^{(0)}=3.422$, $S_{\rm gr}^{(1)}=3.472$ and
$S_{\rm gr}^{(2)}=3.695$ for the WMAP five-year cosmology assumed.
The difference between the dotted red curve and the solid black line shows the
discrepancy between second-order perturbation theory prediction
and $N$-body simulation measurements.
Hence, we find that perturbation theory considerably disagrees with the
numerical simulation result at the smoothing scale of $R_{\rm G}=22h^{-1}$Mpc.
The gravitational
evolution produces a negative shift and decreases the genus amplitude by
$\sim 3\%$.

The top part of Table \ref{tab:sysbias} lists the genus-related statistics
of the samples used in Figure~\ref{fig:sysbias_dm} to quantify the deviations 
of the genus curves from the Gaussian expectation -- 
due to the mentioned systematic effects.

We then consider the effect of galaxy (halo) biasing on the genus -- along 
with shot noise, past light cone gradient and RSD.
To this end, we made 81 mock samples  from the
snapshot halo full cubic data at $z=0$, in exactly the
same way as the past light cone of the LRG mock samples.
We then compared the genus curve averaged over all the halo density fields,
${\langle}G^{\rm j}_{\rm h,r}|{z_{\rm f}}{\rangle}$, where here $h$ stands for halo,
with the one obtained by averaging the 81 matter density fields 
at the same redshift, ${\langle}G^{\rm j}_{\rm m,r}|{z_{\rm f}}{\rangle}$.
Results are displayed in Figure \ref{fig:sysbias_halo}(a), 
with the black solid line for the genus obtained from the averaged 
matter density field, and with the dashed blue line for the one obtained from
an average of the halo density field in real space.
Similarly as in Figure \ref{fig:sysbias_dm}, the middle part in the same panel
displays the quantity $\Delta$ previously defined,
which proves that
the dark matter density field and the halo number
density have very different topology at $22\hMpc$ smoothing scale.
Note however that
the genus curve here includes shot noise
due to discrete sampling of the galaxy density field
as well as the galaxy biasing effect.
Their combined effect has been presented by Hikage, Taruya, \& Suto (2001, 2003)
and Park, Kim, \& Gott (2005a).
As it can be seen from the scatter between the two curves,
the combined effect of galaxy biasing and shot noise
yields significantly larger non-Gaussianities
than the nonlinear gravitational evolution effect.
The bottom part of the same figure shows the combined effect.
In particular, the combined effect increases the genus amplitude and strongly
alters the skewness of the genus curve
(see the value of ${\langle}G^{\rm j}_{\rm h,r}|{z_{\rm f}}{\rangle}$ 
in Table \ref{tab:sysbias}; shifted towards meatball topology,
more percolated and thus larger void structures, and larger number of isolated
clusters compared to those from the matter density fields, 
${\langle}G^{\rm j}_{\rm m,r}|{z_{\rm f}}{\rangle}$).

Past light cone effects and RSDs are quantified in panel (b) of 
Figure \ref{fig:sysbias_halo}.
This is achieved by comparing the genus curve measured from 
our 81 past light cone (LC) mock samples in real space,
${\langle}G^{\rm j}_{\rm h,r}|{\rm LC}{\rangle}$ constructed as explained 
in Section \ref{sec:mock},
with the averaged one obtained from the  halo density fields,
${\langle}G^{\rm j}_{\rm h,r}|{z_{\rm f}}{\rangle}$.
Similarly, we can quantify the effect of RSDs by computing the difference between
${\langle}G^{\rm j}_{\rm h,r}|{\rm LC}{\rangle}$  and the average genus curve
measured from the 81 past light cone mock samples but now
in redshift space, ${\langle}G^{\rm j}_{\rm h,z}|{\rm LC}{\rangle}$ -- 
see also Table \ref{tab:sysbias}.
All these curves are displayed in Figure \ref{fig:sysbias_halo}. 
Again, the lower parts in the same panel clearly describes these effects: 
basically,
the systematic effect introduced by the past light cone on the genus curve 
reaches up to $2\%$ (see the black solid line in the bottom part) 
which is nearly as significant as the gravitational 
evolution effect -- see the black solid line in the middle part of
Fig.~\ref{fig:sysbias_halo}(a) --, 
while RSDs make the amplitude of the genus curve
decrease, but do not alter its shape significantly.

\begin{figure*}
\epsscale{0.8}
\plotone{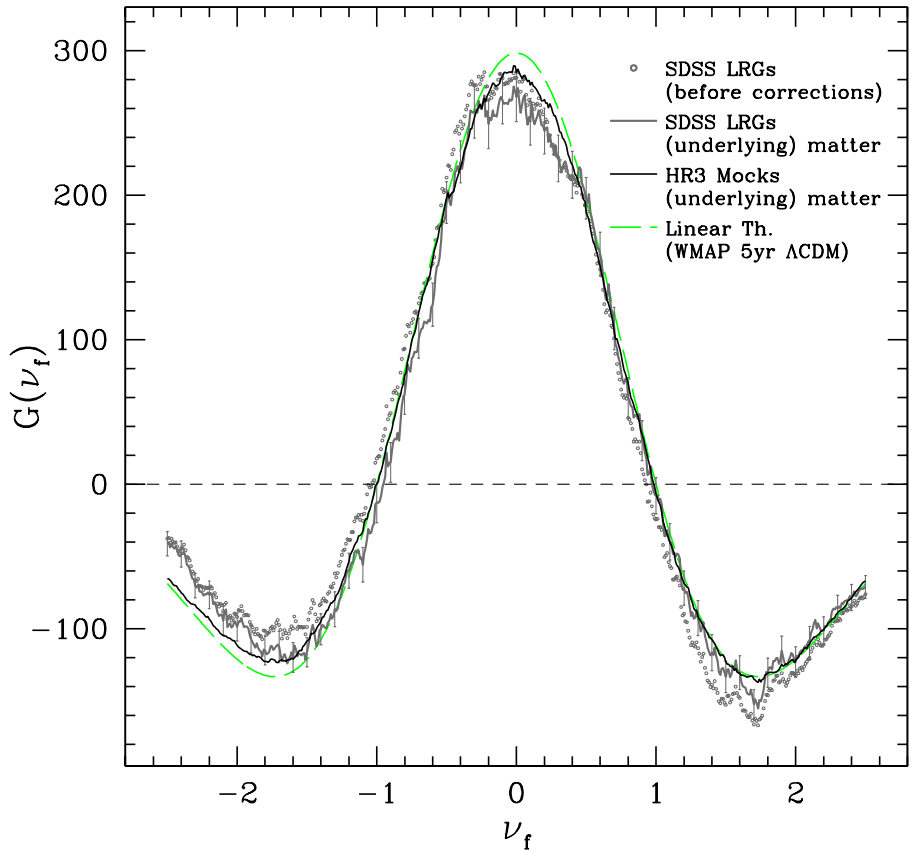}
\caption{
 Observed genus curve before the corrections for systematics (open circles),
   and after applying the corrections (grey thick solid line, 
   with 1$\sigma$ error bars). For comparison, the black thin solid line is
   the genus curve averaged over the 81 mock surveys with the same correction applied,
   and the green dashed line is the linear theory prediction. 
    Overall, the shape of the SDSS LRG genus curve agrees very well
    with the mean topology of the HR3 mock surveys in the $\Lambda$CDM universe.
}
\label{fig:gplot}
\end{figure*}

With the help of our mocks, we are able to
understand and quantify the nonlinear systematics involved in the observational
sample. In particular, we
find that a considerable portion of nonlinearity comes from
the combined effects of galaxy biasing and shot noise.
We can remove the nonlinearities due to those systematics, and
thus estimate the
genus curve of the underlying matter density field from the observed sample.
The correction term that should apply to the observed data, 
derived from the $j$-th mock sample, is given by
\bea
\Delta G^{\rm j}_{\rm sys}=G^{\rm j}_{\rm h,z}|{\rm LC}-G_{\rm m,r}|z_{\rm f}.
\eea
This correction removes all the systematic effects previously mentioned, including RSD, survey mask, shot noise and galaxy biasing,
from the observed genus curve $G_{\rm o}$; clearly, the underlying assumption
is that our HGC assignment scheme is able to model correctly the relation between galaxies in our sample
and the underlying halos.
The genus of the observed (underlying) matter density
field at $z=0$ in real space, reconstructed by applying the correction term, is calculated as follows:
\begin{equation}
G_{\rm o,m,r}=G_{\rm o}-\Delta G_{\rm pix}-\displaystyle\sum\limits_{\rm j=1}^{\rm N}
(\Delta G^{\rm j}_{\rm sys}-\Delta G^{\rm j}_{\rm sim})/N,
\end{equation}
where now $N$ is the number of mock surveys.
We also included the correction for
finite pixel size effect, $\Delta G_{\rm pix}$, and one for the bias
of the $j$-th mock sample 
originated from the cosmic variation in the initial conditions of the HR3 
simulation, $\Delta G^{\rm j}_{\rm sim}=G^{\rm j}_{\rm m,r}|z_{\rm i}-G_{\rm LIN}$.
The correction for the finite pixel size effect is given by the following 
equation:
\bea
&&\Delta G_{\rm pix}=A~{\rm exp}[{\nu_{\rm f}^2/2}]\times\nonumber \\
&&[aH_{0}+bH_{1}(\nu_{\rm f})+cH_{2}(\nu_{\rm f})+dH_{4}(\nu_{\rm f})]{p^2/R_{\rm G}^2},
\label{Eq:pix}
\eea
where $A$ is the genus amplitude calculated from the variance and derivative
(i.e. $\sigma_0$, $\sigma_1$) of the matter density
field on a mesh with vanishing pixel effect (where $p\simeq 0$),
and the coefficients
of each Hermite polynomial, $a$, $b$, $c$, and $d$, are
0.04794, 0.02337, 0.33146, and 0.03843, respectively.
For the density field smoothed with a Gaussian filter
$R_{\rm G}=22\hMpc$ and pixel size of $p=3.7\hMpc$, this effect
can be as large as about $1\%$, and should be taken into account.
Full details of the modeling of the pixel effect can be
found in Young-Rae Kim et al. (in preparation).


\subsection{Genus curve after corrections for systematics} \label{sec:sys}

\begin{figure}
\epsscale{1.}
\plotone{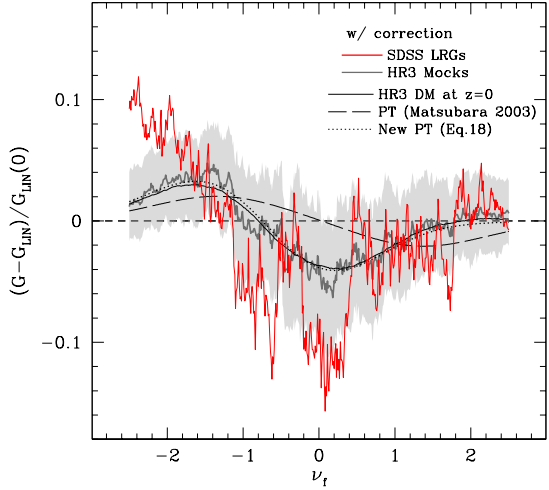}
\caption{
Difference between the genus curves (denoted by $G$) of the underlying
matter distribution derived from the SDSS LRG distribution (thick red line)   
after the correction for systematic effects
and the HR3 snapshot matter distribution at $z=0$ in real space (black thin
line), and the linear analytical prediction for the WMAP 5-year cosmology 
(i.e. $G_{\rm LN}$)
normalized by the amplitude from the linear prediction. 
The shaded area indicates the 1$\sigma$ limits calculated
   from the 81 HR3 halo mock surveys after the same correction has been applied.
   The thin solid gray line shows the averaged deviation from the mock surveys.
The dashed line shows
the second-order perturbation expectation at the median redshift $z=0.28668$ 
of the SDSS sample given by Matsubara (2003); clearly,  the underlying 
matter distribution at the present epoch needs additional terms, 
compared to the second-order perturbation expectation.
}
\label{fig:gplotdiff}
\end{figure}

\begin{figure}
\epsscale{1.5}
\plotone{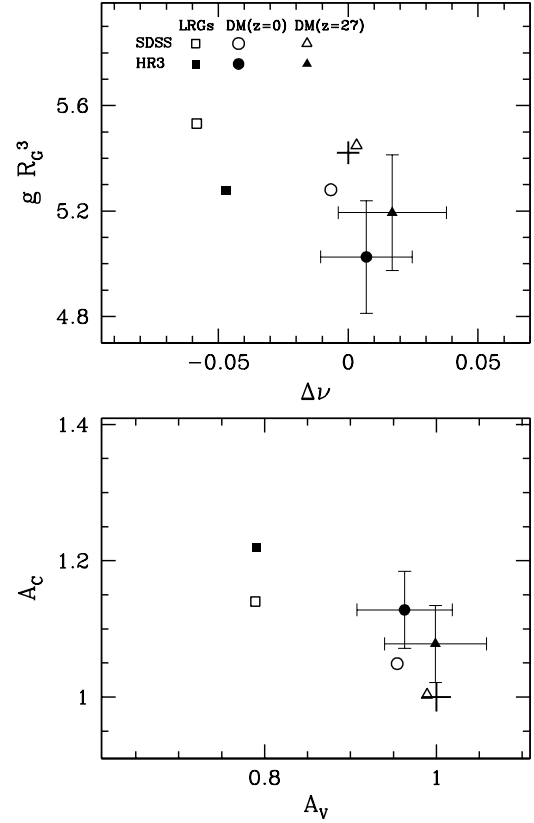}
\caption{Genus-related statistics for the genus curves in Fig.~\ref{fig:gplot}.
Thick crosses are the statistics for the random phase fluctuations.
Squares and circles are the distributions of the LRG density and 
derived matter density, respectively. 
Triangles are statistics from the real space genus curve of the
dark matter particle distribution at the initial epoch
that includes only the contribution of primordial non-Gaussianity.
Filled and open symbols represent the observed and simulated cases, respectively.
See the main text for more details.
}
\label{fig:gpara}
\end{figure}

Finally, we are able to obtain the genus curve of the underlying matter density 
field from the SDSS LRG sample with the effects of shot noise, 
galaxy bias, RSDs, survey boundary and finite pixel size all corrected.
Our final result contains only the
nonlinearity produced by nonlinear gravitational evolution, and 
a possible primordial non-Gaussian component -- if any. 
 Figure~\ref{fig:gplot} shows the observed genus curve 
   before the corrections for systematics (open circles), and after applying 
   those corrections (grey thick solid line, with 1$\sigma$ error bars).
   In the same figure, we also display the genus curve averaged over all the 
   mock surveys after applying the same systematics correction (black thin solid line),
   and the linear theory prediction (green dashed line). 
Overall,
the shape of the genus curve agrees very well with the mean topology of the 
SDSS LRG mock surveys in the $\Lambda$CDM Universe (see also the appendix for 
 the tables of the genus curves, and for more details on the effect of these corrections on the mocks).
However, comparison with simulations also shows small deviations of the 
observed genus curve from the theoretical expectation for Gaussian initial 
conditions.
   Figure \ref{fig:gplotdiff} quantifies these deviations, by showing
   the difference between the genus curves for the SDSS (thick red line) --
   after the correction for systematic effects --  and
   the simulated dark matter distribution at $z=0$  (thin black line),
   and the linear analytical predictions $G_{\rm LN}$ 
   (normalized by the maximum value of $G_{\rm LN}$).
   The shaded area indicates the 1$\sigma$ limits calculated
   from the 81 HR3 halo mock surveys, after the same correction has been applied.
   The thin solid gray line shows the averaged deviation from the mock surveys.
Taking those noisy level-to-level variations into account, and given the
uncertainties, the observed underlying matter distribution is in 
good agreement with the genus computed from the $N$-body simulation and the 
HGC galaxy formation model
-- except for the under-dense regions below $\nu_{\rm f}=-1.8$
filling 3.5\% of the sample volume.
The dashed  line shows
the second-order perturbation expectation at the median redshift $z=0.28668$ 
of the SDSS sample given by Matsubara (2003).
However, the genus curve measured from the gravitationally evolved matter 
density field in the HR3 simulation (solid line in Fig.~\ref{fig:gplotdiff}) 
indicates that perturbation theory cannot model 
properly  gravitational evolution effects.
On the contrary, those effects are modeled well if we add extra terms 
in the second-order perturbation formula of Matsubara (2003), 
depending on $H_0$ and $H_2$.
To this end, the dotted line is a fit to the simulation using the following new 
perturbation formula:
\begin{eqnarray}
\label{eq:new_pert}
G^{\rm{NG}}(\nu_{\rm f})&=&-G(0)~{\rm exp}[{-\nu_{\rm f}^2 /2}]\times\nonumber\\
&&\{\sigma_{0}\left[(S_{\rm gr}^{(1)}-S_{\rm gr}^{(0)})H_3(\nu_{\rm f})+
(S_{\rm gr}^{(2)}-S_{\rm gr}^{(0)})H_1(\nu_{\rm f})\right]\nonumber\\
&&+\sigma_{0}^2\left[A_0 H_0(\nu_f)+A_2H_2(\nu_{\rm f})\right]\},
\end{eqnarray}
where $G(0)=298.5$.
The second-order perturbation theory predicts
the power spectrum $P(k,z)$ in Equation~\ref{eq:sigma} as follows: 
$P(k,z)=P_{\rm LIN}(k,z)W^2(kR)+{\it O}(\sigma_0^4)$,
where $P_{\rm LIN}(k,z)$ is the linear power spectrum and
$\sigma_0$ (up to the lowest order) is 0.15381 at the median redshift $z=0.28668$
of the SDSS sample.
The additional coefficients $A_{0}$ and $A_{2}$ 
are $-0.2226$ and $1.4702$, respectively.
Table~\ref{tab:gplot} lists the genus-related statistics for
all the genus measurements relative to Figure \ref{fig:gplotdiff}.

\begin{table*}
\caption{Genus-related statistics for the genus curves in Figure~\ref{fig:gpara}.}
\begin{center}
\begin{tabular}{lllll}
\hline\hline
Genus                    &  $G_{\rm fit}$ & $\Delta \nu_{\rm f}$  & $A_{\rm V}$ & $A_{\rm C}$ \\ \hline
Observation &&&& \\\hline
$G_{\rm o}$              &  285.2   & -0.047 &  0.79 & 1.22 \\
$G_{\rm o,m,r}|z=0$   &  271.7   &  0.007 &  0.96 & 1.13 \\
$G_{\rm o,m,r}|z_{\rm pri}$ &  280.8   &  0.017 &  1.00 & 1.08 \\  \hline
Simulation &&&& \\\hline
$G_{\rm M}$     &$299.0\pm 11.5$ & $-0.058 \pm 0.018$ & $0.79\pm0.06$ & $1.14\pm 0.06$\\
$G_{{\rm M,m},r}|z=0 $ & $285.5$ & $-0.007$ & $0.95$ & $1.05$\\
$G_{{\rm M,m},r}|z_{\rm pri}$   & $294.5$ & $ 0.003$ & $0.99$ & $1.00$\\
\hline
\end{tabular}\end{center}
{\bf Notes.} $G_{\rm o}$ is the genus value of the SDSS LRG sample, and
$G_{\rm M}$ the one averaged over all the 81 light cone LRG mock samples,
$\langle G^{\rm j}_{\rm h,z}|{\rm LC}\rangle$.
$G_{\rm o,m,r}|z=0$ and $G_{\rm M,m,r}|z=0$ are genus values
of the underlying dark matter distributions derived from $G_{\rm o}$ and $G_{\rm M}$, respectively --
after corrections for systematics. $G_{\rm o,m,r}|z_{\rm pri}$
and $G_{\rm M,m,r}|z_{\rm pri}$ are real space genus values of the 
dark matter distribution at the initial epoch of the simulation including only contribution
of primordial non-Gaussianity, for both the observation and simulations.
\label{tab:gplot}
\vspace{1cm}
\end{table*}

Finally, in Figure \ref{fig:gpara} we present
the genus-related statistics for the previous genus curves; this is 
helpful in order to understand systematic biases.
The statistics for the random phase fluctuations are indicated
by thick crosses.
Triangles are statistics from the genus curve of the derived real-space
  dark matter particle distribution at the initial epoch of the simulation
  ($G_{{\rm M,m,r}}|{z_{\rm pri}}$ and $G_{{\rm o,m,r}}|{z_{\rm pri}}$),
  which includes only a primordial non-Gaussianity contribution for 
  both the observation and simulation.
  The correction applied here has only the contribution of non-Gaussianity
  produced by nonlinear gravitational evolution,
  ${\langle}G^{j}_{{\rm m},r}|{z_{\rm i}}{\rangle}-
  {\langle}G^{j}_{{\rm m},r}|{z_{\rm f}}{\rangle}$ in Table \ref{tab:sysbias}.
The distribution of the matter density (circles) has a smaller overall amplitude of
the genus curve, more voids, and fewer clusters, and is bubble
shifted compared to that of the galaxy density (squares).
Triangles  are obtained from the real space genus curve of the dark matter particle 
distribution at the initial epoch of the simulation, which
includes only a primordial non-Gaussian contribution.
The difference between
circles and triangles indicates the effect of nonlinear
gravitational evolution.
While the statistics of the initial matter density field of the $\Lambda$CDM
  $N$-body simulation with primordial Gaussianity (open triangles)
  are nearly the same as those of the 
  random phase curve (as expected), 
  the statistics of the initial matter density field obtained from the LRG sample (filled triangles) 
  show still deviations from the Gaussian expectation;
  $A_{\rm V}$ and $\Delta \nu_{\rm f}$ are within about $1\sigma$ of the 
  Gaussian values, and the amplitude and $A_{\rm C}$ show a relatively 
  large difference between the observation and the Gaussian prediction.



\section{Conclusions} \label{sec:summary}
In this paper we presented measurements of the genus topology of LRGs
from the SDSS DR7 catalog, with unprecedented
statistical significance.
We made a volume-limited sample in the redshift range $0.16 < z <  0.36$ and
and rest-frame $\rm g$-band absolute magnitudes of $-23.2<M_{\rm g}<-21.1$
using the DR7-Full sample of Kazin et al. (2010) and then
imposed some additional cuts as in Choi et al. (2010), 
which leave a total of $60,466$ LRGs over about 2.33 sr.
We constructed the galaxy mass and number density fields 
from the observed LRG sample, using a novel technique -- 
called `luminosity function matching' -- outlined in 
Section \ref{sec:data_methods}, and computed the observed genus curve.
We also produced 81 independent mock LRG samples from the HR3 (Kim et al. 2011),
one of the largest $N$-body simulations currently available,
that evolved $7210^3$ particles in a $10815 h^{-1}$Mpc size box.
The construction of simulated LRG catalogs required several subtle steps,
explained in Section \ref{sec:sims_methods}. In particular, we adopted the halo-galaxy one-to-one monotonic correspondence model (HGC) of Kim, Park, \& Choi (2008)
to populate dark matter halos with galaxies, and identified LRGs as the most massive subhalos.

Thanks to the unprecedented volume of the HR3, we were able to
carefully model and study all the known systematics which affect the genus curve, 
such as finite pixel size, survey boundary mask, radial and angular selection
function, past light cone gradient, initial conditions of the simulations,
shot noise, cosmic variance, RSDs and galaxy biasing.
Upon removal of all known systematics,
our final genus curve (Section \ref{sec:sys}, Figure \ref{fig:gplot})
contains only the nonlinearity produced by
nonlinear gravitational evolution, and a possible 
primordial non-Gaussian component -- if any.
In particular, we find the observed genus amplitude
to reach 285 with an uncertainty of 4.0\% including cosmic variance
(before the correction for the systematics): this is
the most accurate constraint on the genus amplitude to date,
which significantly improves on our previous measurement (Gott et al. 2009).
Overall, the shape of the observed genus curve agrees very well with the 
mean topology of the SDSS LRG mock surveys in the $\Lambda$CDM universe, and 
this should be considered as a success of our large volume $N$-body simulation, 
as well as of our procedure to construct mock LRG samples from the HR3
(see Fig.~\ref{fig:main_result}).

However, comparison with simulations also shows small but significant 
deviations of the observed genus curve from the theoretical expectation 
for Gaussian initial conditions:
Figures \ref{fig:gplotdiff}  and \ref{fig:gpara}
show explicitly these deviations.
We used genus-derived statistics (Section \ref{sec:grs}, Section \ref{sec:ngd} 
and Tables 1-3) to
estimate and quantify departures from Gaussianity of the genus curve.
While a consistent
part of the non-Gaussian deviations is caused by systematics, and mainly driven 
by shot noise and biasing, 
removing their effects on the genus curve still leaves some discrepancies 
from the Gaussian expectations.
This fact can be attributed to the nonlinearity produced by
gravitational evolution, in addition to a possible primordial 
non-Gaussian component.
We investigated here the role of nonlinear gravitational evolution 
on the genus curve, while in a forthcoming publication we will
provide an interpretation of the remaining deviations in the context of 
primordial non-Gaussianity.
In particular, in this study we found that the second-order perturbation theory 
prediction of Matsubara (2003) disagrees significantly with
the genus curve measured from the gravitationally evolved matter density field 
in the HR3 simulation (solid line in Fig.~\ref{fig:gplotdiff}).
On the contrary, the nonlinear gravitational evolution effects 
are modeled well if we add extra terms in the second-order perturbation 
formula of Matsubara; 
to this end, we also provided a new second-order perturbation formula 
(Eq.~\ref{eq:new_pert}) which better fits our results.

In summary, the main achievements of this paper are as follows:
\begin{enumerate}[i.]
\item  
We measured the genus amplitude from the SDSS LRG volume-limited sample, 
and found it to reach 285 with an uncertainty of 4.0\% including cosmic 
variance; this is the most accurate constraint on the genus amplitude to date,
which significantly improves on the results by Gott et al. (2009).
\item  
The overall shape of the observed genus curve agrees very well with 
the mean topology of the SDSS LRG mock surveys in the $\Lambda$CDM universe,
confirming the correctness of our large volume $N$-body simulation and 
procedure to construct mock LRG samples.
This should also be seen as another strong support of the $\Lambda$CDM paradigm, 
similar to the one recently presented by Park et al. (2012), 
who was able to prove that observed high- and low-density LSSs have the 
richness/volume and size distributions consistent with the $\Lambda$CDM universe.
\item 
Thanks to our unprecedented large volume simulation (HR3), 
we gained an excellent control of the numerous systematics which affect 
the genus curve, ranging from observational to statistical or cosmological 
effects which introduce non-Gaussianities in the genus shape. 
We were able to successfully model and remove all the known systematics.
\item 
We proposed a new method to
construct the galaxy mass and number density fields from the observed 
LRG sample (i.e. the `luminosity function matching' technique), and an 
accurate procedure to construct LRG mock samples.
\item 
We proved that second-order perturbation theory 
(Hikage et al. 2002; Matsubara 2003) cannot model
the genus curve measured from the gravitationally evolved matter density field 
in the HR3 simulation, and provided a new fitting formula
which adds extra terms to the original second-order perturbation expression 
and matches better our results.
\end{enumerate}

After removing all known systematics and modeling the nonlinear gravitational 
evolution more accurately, we are still left with some additional 
non-Gaussian signal.
Clearly, since we were able for the first time to isolate and quantify a 
non-Gaussian contribution directly
from the observed genus curve which is not due to systematics, and 
argued that even an improved second-order perturbation theory cannot explain 
all the non-Gaussian discrepancies, our next step is to interpret 
those deviations in the context of the local  $f_{\rm NL}$-type model 
(Salopek \& Bond 1990; Gangui et al. 1994; Verde et al. 2000; 
Komatsu \& Spergel 2001).
This will allow us to constrain the standard non-Gaussianity parameter $f_{\rm NL}$ 
directly from topological measurements of the LSS.



\begin{acknowledgements}

We would like to warmly thank Changbom Park for his supervision, help and constant feedback 
throughout the entire project.
This work was supported by a grant from Kyung
Hee University in 2011 (KHU-20100179). 
S.S.K. was supported by a Mid-career Reaserch Program (No. 2011-0016898) through
the National Research Foundation (NRF) grant funded by the Ministry
of Education, Science and Technology (MEST) of Korea.
Support for this work was also provided by the National Research Foundation of Korea to
the Center for Galaxy Evolution Research (No. 2010-0027910).
We thank the
Korea Institute for Advanced Study for providing computing
resources (KIAS Center for Advanced Computation
Linux Cluster System) for this work.

Funding for the SDSS and SDSS-II has been provided by
the Alfred P. Sloan Foundation, the Participating Institutions,
the National Science Foundation, the U.S. Department of
Energy, the National Aeronautics and Space Administration,
the Japanese Monbukagakusho, the Max Planck Society, and
the Higher Education Funding Council for England. The SDSS
Web site is http://www.sdss.org/.

The SDSS is managed by the Astrophysical Research Consortium
for the Participating Institutions. The Participating
Institutions are the American Museum of Natural History,
Astrophysical Institute Potsdam, University of Basel, Cambridge
University, Case Western Reserve University, University
of Chicago, Drexel University, Fermilab, the Institute
for Advanced Study, the Japan Participation Group, Johns
Hopkins University, the Joint Institute for Nuclear Astrophysics,
the Kavli Institute for Particle Astrophysics and Cosmology,
the Korean Scientist Group, the Chinese Academy of Sciences
(LAMOST), Los Alamos National Laboratory, the Max-Planck-Institute
for Astronomy (MPIA), the Max-Planck-Institute for
Astrophysics (MPA), New Mexico State University, Ohio State
University, University of Pittsburgh, University of Portsmouth,
Princeton University, the United States Naval Observatory, and
the University of Washington.
\end{acknowledgements}



\appendix


\section{Tables of genus curves} \label{sec:genus_curves}

In this appendix we provide tables of the genus curves for
the reader who wish to use galaxy clustering
topology.
Table~\ref{tab:mgplot_genus} contains the
mean genus values measured from number density fields 
and mass-weighted density fields of the 81 past light cone mock samples,
and the genus values of the SDSS LRGs,
as a function of both volume-fraction threshold level ($\nu_{\rm f}$) and direct
density threshold level ($\nu$) -- as measured in
Sections~\ref{sec:weight} and in Appendix~\ref{sec:dthres}; see also the
corresponding genus curves, plotted in 
Figures~\ref{fig:mgplot} and \ref{fig:gplot_dth}.
We additionally provide similar quantities for 
the dark matter particle distribution in the $\Lambda$CDM model
at the current epoch, $G_{\rm m,r}|z=z_{\rm f}$, for comparison.
The observed genus values ($G_{\rm o}$ and $G_{\rm o,m,r}$) before 
and after the correction for systematic effects as a function
of volume-fraction threshold levels, plotted in
Figure~\ref{fig:gplot}, are listed in Table~\ref{tab:gplot_matter}.
$G_{\rm M}$ and $G_{\rm M,m,r}$ are the mean genus values before
and after the applied corrections, measured from
the 81 light cone mock LRG samples in the same way as done for the SDSS sample.
Electronic forms of
these tables are available from the authors upon request.
\begin{table}
\caption{Genus values at a given threshold level of the samples used in 
Figures~\ref{fig:mgplot} and \ref{fig:gplot_dth}}
\begin{center}
\begin{tabular}{r|rr|rr|rr}
\hline\hline
&\multicolumn{4}{c|}{Volume Fraction, $\nu_{\rm f}$} &\multicolumn{2}{c}{Direct Density, $\nu$}\\
\cline{2-5} \cline{6-7}
&\multicolumn{2}{c}{Mock LRGs} &\multicolumn{2}{c|}{SDSS LRGs}&\multicolumn{1}{c}{Mock LRGs} &\multicolumn{1}{c}{SDSS LRGs}\\
\cline{2-3} \cline{4-5} \cline{6-7}
$\nu_{\rm f}$, $\nu$& Number & Mass & Number & Mass &  Number & Number \\
\hline
$ -2.5$&$ -61.7\pm   8.4$& $  -58.6\pm   8.6$& $  -37.6$&$  -46.0$ & $  -0.0\pm   0.0$&$    0.0$ \\
$ -2.4$&$ -71.1\pm   9.1$& $  -67.9\pm   9.3$& $  -44.2$&$  -57.0$ & $  -0.0\pm   0.2$&$    0.0$ \\
$ -2.3$&$ -80.1\pm   8.6$& $  -76.7\pm   8.5$& $  -66.6$&$  -63.5$ & $  -0.6\pm   0.8$&$   -2.0$ \\
$ -2.2$&$ -88.7\pm   9.6$& $  -84.3\pm   8.9$& $  -71.5$&$  -65.6$ & $  -3.2\pm   1.9$&$   -3.0$ \\
$ -2.1$&$ -98.4\pm  10.0$& $  -92.6\pm   9.9$& $  -87.9$&$  -71.4$ & $  -9.9\pm   3.4$&$   -8.9$ \\
$ -2.0$&$-104.6\pm   9.7$& $  -97.8\pm   9.3$& $  -91.9$&$  -88.8$ & $ -22.0\pm   5.1$&$  -20.8$ \\
$ -1.9$&$-108.1\pm   9.8$& $ -101.8\pm   9.5$& $  -97.5$&$  -86.9$ & $ -40.3\pm   6.9$&$  -30.3$ \\
$ -1.8$&$-108.9\pm  10.0$& $ -103.1\pm  11.6$& $ -101.4$&$ -103.3$ & $ -62.5\pm   7.6$&$  -41.9$ \\
$ -1.7$&$-107.1\pm  10.7$& $  -99.4\pm  11.8$& $  -97.4$&$ -100.9$ & $ -84.4\pm   8.3$&$  -70.9$ \\
$ -1.6$&$-100.9\pm  10.2$& $  -93.5\pm  10.1$& $  -99.7$&$  -91.5$ & $-102.0\pm   8.4$&$  -91.6$ \\
$ -1.5$&$ -91.1\pm  10.2$& $  -83.2\pm   9.7$& $  -98.7$&$  -83.5$ & $-109.2\pm  10.6$&$ -100.5$ \\
$ -1.4$&$ -77.3\pm  10.8$& $  -68.1\pm  10.5$& $  -82.9$&$  -67.1$ & $-105.9\pm  11.3$&$ -104.4$ \\
$ -1.3$&$ -55.7\pm  12.1$& $  -48.5\pm  11.9$& $  -61.7$&$  -55.0$ & $ -90.9\pm  11.6$&$  -98.6$ \\
$ -1.2$&$ -31.8\pm  12.1$& $  -23.5\pm  12.2$& $  -38.5$&$  -31.1$ & $ -63.0\pm  11.8$&$  -76.9$ \\
$ -1.1$&$  -2.3\pm  11.9$& $    4.6\pm  12.8$& $  -25.4$&$  -22.8$ & $ -26.6\pm  12.3$&$  -38.0$ \\
$ -1.0$&$  30.4\pm  14.0$& $   35.1\pm  11.5$& $   17.8$&$   11.9$ & $  18.0\pm  13.6$&$   -3.1$ \\
$ -0.9$&$  68.9\pm  13.7$& $   69.8\pm  12.3$& $   49.3$&$   61.5$ & $  70.7\pm  14.3$&$   48.8$ \\
$ -0.8$&$ 106.5\pm  13.8$& $  107.6\pm  13.1$& $   93.6$&$   89.6$ & $ 118.7\pm  13.6$&$  110.3$ \\
$ -0.7$&$ 143.6\pm  13.7$& $  144.5\pm  14.2$& $  125.3$&$  106.6$ & $ 167.0\pm  14.3$&$  140.0$ \\
$ -0.6$&$ 183.5\pm  15.0$& $  179.2\pm  14.8$& $  152.6$&$  155.9$ & $ 209.8\pm  13.8$&$  201.0$ \\
$ -0.5$&$ 215.2\pm  14.4$& $  210.4\pm  15.2$& $  212.1$&$  205.2$ & $ 246.6\pm  15.4$&$  249.0$ \\
$ -0.4$&$ 247.2\pm  14.5$& $  238.7\pm  16.6$& $  250.9$&$  249.2$ & $ 273.9\pm  18.7$&$  273.8$ \\
$ -0.3$&$ 271.5\pm  18.1$& $  260.2\pm  15.7$& $  275.8$&$  276.6$ & $ 289.7\pm  17.3$&$  269.1$ \\
$ -0.2$&$ 287.2\pm  16.8$& $  276.3\pm  16.7$& $  265.8$&$  282.9$ & $ 299.2\pm  15.3$&$  277.6$ \\
$ -0.1$&$ 297.5\pm  16.3$& $  282.5\pm  14.7$& $  280.9$&$  278.4$ & $ 296.5\pm  15.7$&$  281.0$ \\
$  0.0$&$ 298.2\pm  17.1$& $  280.0\pm  16.4$& $  281.1$&$  272.4$ & $ 284.6\pm  17.0$&$  262.5$ \\
$  0.1$&$ 289.9\pm  16.1$& $  269.8\pm  16.3$& $  264.5$&$  280.0$ & $ 264.8\pm  16.1$&$  243.5$ \\
$  0.2$&$ 271.6\pm  16.0$& $  252.6\pm  14.8$& $  250.0$&$  259.4$ & $ 241.0\pm  17.5$&$  216.8$ \\
$  0.3$&$ 248.1\pm  16.2$& $  230.3\pm  15.7$& $  225.9$&$  228.5$ & $ 209.9\pm  17.5$&$  208.0$ \\
$  0.4$&$ 218.0\pm  16.3$& $  197.2\pm  16.8$& $  211.1$&$  197.2$ & $ 178.6\pm  14.8$&$  192.6$ \\
$  0.5$&$ 183.0\pm  14.4$& $  161.1\pm  15.3$& $  193.0$&$  163.4$ & $ 140.4\pm  15.7$&$  144.2$ \\
$  0.6$&$ 140.5\pm  15.2$& $  120.7\pm  12.2$& $  148.6$&$  143.9$ & $ 102.9\pm  17.1$&$  102.2$ \\
$  0.7$&$  98.0\pm  14.7$& $   79.2\pm  13.6$& $  100.9$&$  111.9$ & $  68.2\pm  16.4$&$   61.4$ \\
$  0.8$&$  55.8\pm  14.1$& $   38.2\pm  13.2$& $   45.9$&$   58.8$ & $  32.2\pm  15.4$&$   27.1$ \\
$  0.9$&$  14.5\pm  13.9$& $   -1.0\pm  12.8$& $   10.3$&$   -1.9$ & $   1.3\pm  15.5$&$  -12.1$ \\
$  1.0$&$ -23.0\pm  12.5$& $  -38.2\pm  12.2$& $  -24.5$&$  -39.0$ & $ -30.6\pm  14.4$&$  -27.1$ \\
$  1.1$&$ -56.9\pm  12.7$& $  -68.1\pm  11.2$& $  -56.6$&$  -88.6$ & $ -57.0\pm  14.7$&$  -60.6$ \\
$  1.2$&$ -86.6\pm  11.8$& $  -94.3\pm  11.8$& $  -89.6$&$ -120.9$ & $ -80.7\pm  13.8$&$  -87.1$ \\
$  1.3$&$-112.3\pm  12.2$& $ -115.9\pm  10.8$& $ -113.5$&$ -149.6$ & $-100.6\pm  13.2$&$ -112.4$ \\
$  1.4$&$-129.7\pm  12.5$& $ -129.6\pm  10.8$& $ -138.8$&$ -162.3$ & $-117.7\pm  12.6$&$ -116.1$ \\
$  1.5$&$-142.4\pm  11.0$& $ -142.2\pm  11.4$& $ -152.8$&$ -145.0$ & $-130.2\pm  12.7$&$ -142.0$ \\
$  1.6$&$-148.2\pm  11.8$& $ -147.4\pm  11.0$& $ -147.9$&$ -169.8$ & $-139.4\pm  12.3$&$ -150.8$ \\
$  1.7$&$-148.9\pm  10.1$& $ -148.7\pm  10.3$& $ -166.3$&$ -159.6$ & $-144.3\pm  11.5$&$ -143.3$ \\
$  1.8$&$-146.3\pm  10.1$& $ -145.2\pm   9.5$& $ -146.0$&$ -159.1$ & $-148.5\pm  12.4$&$ -156.1$ \\
$  1.9$&$-140.4\pm  10.1$& $ -137.8\pm   9.9$& $ -130.8$&$ -138.9$ & $-149.0\pm  10.6$&$ -163.4$ \\
$  2.0$&$-131.1\pm  10.7$& $ -128.9\pm  10.2$& $ -132.4$&$ -139.5$ & $-148.3\pm   9.1$&$ -161.8$ \\  
$  2.1$&$-122.8\pm  10.5$& $ -118.6\pm   9.2$& $ -120.9$&$ -126.0$ & $-145.0\pm   9.8$&$ -142.8$ \\  
$  2.2$&$-109.9\pm   9.8$& $ -107.0\pm   8.7$& $ -110.9$&$ -113.0$ & $-140.8\pm   9.6$&$ -137.5$ \\  
$  2.3$&$ -97.6\pm   9.3$& $  -94.5\pm   8.4$& $  -94.5$&$ -103.8$ & $-135.6\pm  10.0$&$ -137.9$ \\  
$  2.4$&$ -85.2\pm   9.1$& $  -82.0\pm   8.2$& $  -84.8$&$  -93.4$ & $-129.8\pm   9.6$&$ -134.1$ \\  
$  2.5$&$ -72.0\pm   8.0$& $  -70.3\pm   7.5$& $  -76.1$&$  -76.4$ & $-123.6\pm   8.5$&$ -124.6$ \\  
\hline
\end{tabular}
\end{center}
{\bf Notes.} 
`Number' and `Mass' stand for number density field and mass weighted density field,
respectively.
\label{tab:mgplot_genus}
\end{table}

\begin{table}
\begin{center}
\caption{Genus values at a given threshold level of the samples used in 
Figures~\ref{fig:gplot} and \ref{fig:gpara}.} \label{tab:gplots}
\centering
\begin{tabular}{r|rrr|rrrr}
\hline\hline
&\multicolumn{3}{c|}{SDSS} &\multicolumn{4}{c}{HR3}\\
$\nu_{\rm f}$ & $G_{\rm o}$ & $G_{\rm o,m,r}|z=0$ & $G_{\rm o,m,r}|z_{\rm pri}$ &
$G_{\rm M}$ & $G_{\rm M,m,r}|z=0$ & $G_{\rm M,m,r}|z_{\rm pri}$
&$G_{\rm m,r}|z=0$\\ 
\hline
-2.5  &  -37.6 &  -41.2&  -45.8 &$  -61.7\pm   8.4$ &     -65.3 &    -69.8& -64.9\\ 
-2.4  &  -44.2 &  -48.3&  -52.9 &$  -71.1\pm   9.1$ &     -75.1 &    -79.7& -74.8\\
-2.3  &  -66.6 &  -70.1&  -74.3 &$  -80.1\pm   8.6$ &     -83.6 &    -87.7& -85.2\\
-2.2  &  -71.5 &  -76.4&  -82.2 &$  -88.7\pm   9.6$ &     -93.6 &    -99.4& -95.5\\
-2.1  &  -87.9 &  -94.3&  -99.6 &$  -98.4\pm  10.0$ &    -104.8 &   -110.1&-105.2\\
-2.0  &  -91.9 &  -98.7& -105.5 &$ -104.6\pm   9.7$ &    -111.5 &   -118.3&-113.4\\
-1.9  &  -97.5 & -107.1& -115.0 &$ -108.1\pm   9.8$ &    -117.7 &   -125.7&-119.5\\
-1.8  & -101.4 & -115.0& -123.6 &$ -108.9\pm  10.0$ &    -122.6 &   -131.1&-123.5\\
-1.7  &  -97.4 & -112.7& -123.8 &$ -107.1\pm  10.7$ &    -122.4 &   -133.6&-124.1\\
-1.6  &  -99.7 & -119.9& -130.7 &$ -100.9\pm  10.2$ &    -121.0 &   -131.8&-120.5\\
-1.5  &  -98.7 & -116.3& -124.1 &$  -91.1\pm  10.2$ &    -108.6 &   -116.4&-112.5\\
-1.4  &  -82.9 & -100.5& -108.2 &$  -77.3\pm  10.8$ &     -94.9 &   -102.6& -99.3\\
-1.3  &  -61.7 &  -86.4&  -97.8 &$  -55.7\pm  12.1$ &     -80.4 &    -91.7& -80.9\\
-1.2  &  -38.5 &  -59.9&  -63.1 &$  -31.8\pm  12.1$ &     -53.2 &    -56.4& -57.6\\
-1.1  &  -25.4 &  -55.7&  -59.6 &$   -2.3\pm  11.9$ &     -32.7 &    -36.5& -29.2\\
-1.0  &   17.8 &  -12.6&  -14.0 &$   30.4\pm  14.0$ &       0.1 &     -1.4&   3.5\\
-0.9  &   49.3 &   15.1&   14.9 &$   68.9\pm  13.7$ &      34.8 &     34.6&  39.7\\
-0.8  &   93.6 &   61.9&   57.0 &$  106.5\pm  13.8$ &      74.8 &     69.9&  78.0\\
-0.7  &  125.3 &  100.8&  103.2 &$  143.6\pm  13.7$ &     119.2 &    121.5& 117.1\\
-0.6  &  152.6 &  124.3&  126.7 &$  183.5\pm  15.0$ &     155.2 &    157.6& 155.5\\
-0.5  &  212.1 &  194.9&  205.4 &$  215.2\pm  14.4$ &     197.9 &    208.4& 191.7\\
-0.4  &  250.9 &  222.9&  225.7 &$  247.2\pm  14.5$ &     219.2 &    221.9& 224.2\\
-0.3  &  275.8 &  257.4&  271.7 &$  271.5\pm  18.1$ &     253.2 &    267.4& 251.0\\
-0.2  &  265.8 &  249.1&  254.2 &$  287.2\pm  16.8$ &     270.6 &    275.7& 271.4\\
-0.1  &  280.9 &  265.5&  271.0 &$  297.5\pm  16.3$ &     282.1 &    287.6& 283.5\\
 0.0  &  281.1 &  268.0&  275.5 &$  298.2\pm  17.1$ &     285.0 &    292.5& 287.3\\
 0.1  &  264.5 &  250.8&  250.0 &$  289.9\pm  16.1$ &     276.1 &    275.4& 282.4\\
 0.2  &  250.0 &  244.4&  248.9 &$  271.6\pm  16.0$ &     266.0 &    270.6& 267.0\\
 0.3  &  225.9 &  225.3&  234.7 &$  248.1\pm  16.2$ &     247.6 &    257.0& 247.7\\
 0.4  &  211.1 &  214.1&  223.4 &$  218.0\pm  16.3$ &     221.0 &    230.3& 220.0\\
 0.5  &  193.0 &  197.8&  212.0 &$  183.0\pm  14.4$ &     187.8 &    202.0& 186.8\\
 0.6  &  148.6 &  159.2&  169.5 &$  140.5\pm  15.2$ &     151.1 &    161.4& 149.6\\
 0.7  &  100.9 &  107.7&  111.8 &$   98.0\pm  14.7$ &     104.8 &    108.9& 110.2\\
 0.8  &   45.9 &   59.8&   64.1 &$   55.8\pm  14.1$ &      69.7 &     74.1&  69.9\\
 0.9  &   10.3 &   30.1&   41.8 &$   14.5\pm  13.9$ &      34.3 &     46.1&  31.0\\
 1.0  &  -24.5 &   -7.9&   -2.9 &$  -23.0\pm  12.5$ &      -6.5 &     -1.5&  -5.7\\
 1.1  &  -56.6 &  -39.9&  -33.7 &$  -56.9\pm  12.7$ &     -40.1 &    -34.0& -39.0\\
 1.2  &  -89.6 &  -71.9&  -69.5 &$  -86.6\pm  11.8$ &     -68.9 &    -66.5& -67.6\\
 1.3  & -113.5 &  -92.7&  -90.7 &$ -112.3\pm  12.2$ &     -91.6 &    -89.5& -91.3\\
 1.4  & -138.8 & -116.6& -114.0 &$ -129.7\pm  12.5$ &    -107.6 &   -104.9&-109.6\\
 1.5  & -152.8 & -133.3& -133.4 &$ -142.4\pm  11.0$ &    -122.9 &   -123.0&-122.5\\
 1.6  & -147.9 & -130.4& -127.8 &$ -148.2\pm  11.8$ &    -130.7 &   -128.1&-130.0\\
 1.7  & -166.3 & -152.2& -152.2 &$ -148.9\pm  10.1$ &    -134.9 &   -134.9&-133.5\\
 1.8  & -146.0 & -130.4& -128.5 &$ -146.3\pm  10.1$ &    -130.7 &   -128.8&-132.5\\
 1.9  & -130.8 & -115.2& -113.0 &$ -140.4\pm  10.1$ &    -124.9 &   -122.7&-128.0\\
 2.0  & -132.4 & -123.1& -122.7 &$ -131.1\pm  10.7$ &    -121.8 &   -121.4&-120.8\\
 2.1  & -120.9 & -108.8& -108.5 &$ -122.8\pm  10.5$ &    -110.8 &   -110.4&-111.6\\
 2.2  & -110.9 & -101.4& -102.9 &$ -109.9\pm   9.8$ &    -100.5 &   -101.9&-101.3\\
 2.3  &  -94.5 &  -85.9&  -86.6 &$  -97.6\pm   9.3$ &     -89.1 &    -89.7& -90.3\\
 2.4  &  -84.8 &  -78.4&  -78.4 &$  -85.2\pm   9.1$ &     -78.9 &    -78.9& -79.3\\
 2.5  &  -76.1 &  -71.1&  -70.6 &$  -72.0\pm   8.0$ &     -66.9 &    -66.5& -68.3\\
\hline
\end{tabular}\end{center}
\label{tab:gplot_matter}
{\bf Notes.}
$G_{\rm o}$ is the genus value of the SDSS LRG sample, and
$G_{\rm M}$ is the one averaged over all the 81 light cone LRG mock samples,
$\langle G^{\rm j}_{\rm h,z}|{\rm LC}\rangle$.
$G_{\rm o,m,r}|z=0$ and $G_{\rm M,m,r}|z=0$ are the genus values
of the underlying dark matter distribution, derived from $G_{\rm o}$ and $G_{\rm M}$, respectively,
after correcting for the various systematics. $G_{\rm o,m,r}|z_{\rm pri}$
and $G_{\rm M,m,r}|z_{\rm pri}$ are the real space genus values of the
dark matter distribution at the initial epoch of the simulation, which include only the contribution
 of primordial non-Gaussianity (both for the observation and simulation).
$G_{\rm m,r}|z_{\rm f}$ is the genus value for matter distribution in
real space at $z=0$, using the full cubic data.
\end{table}


\section{Threshold Density versus Volume Fraction}\label{sec:dthres}

\begin{figure}
\epsscale{0.8}
\plotone{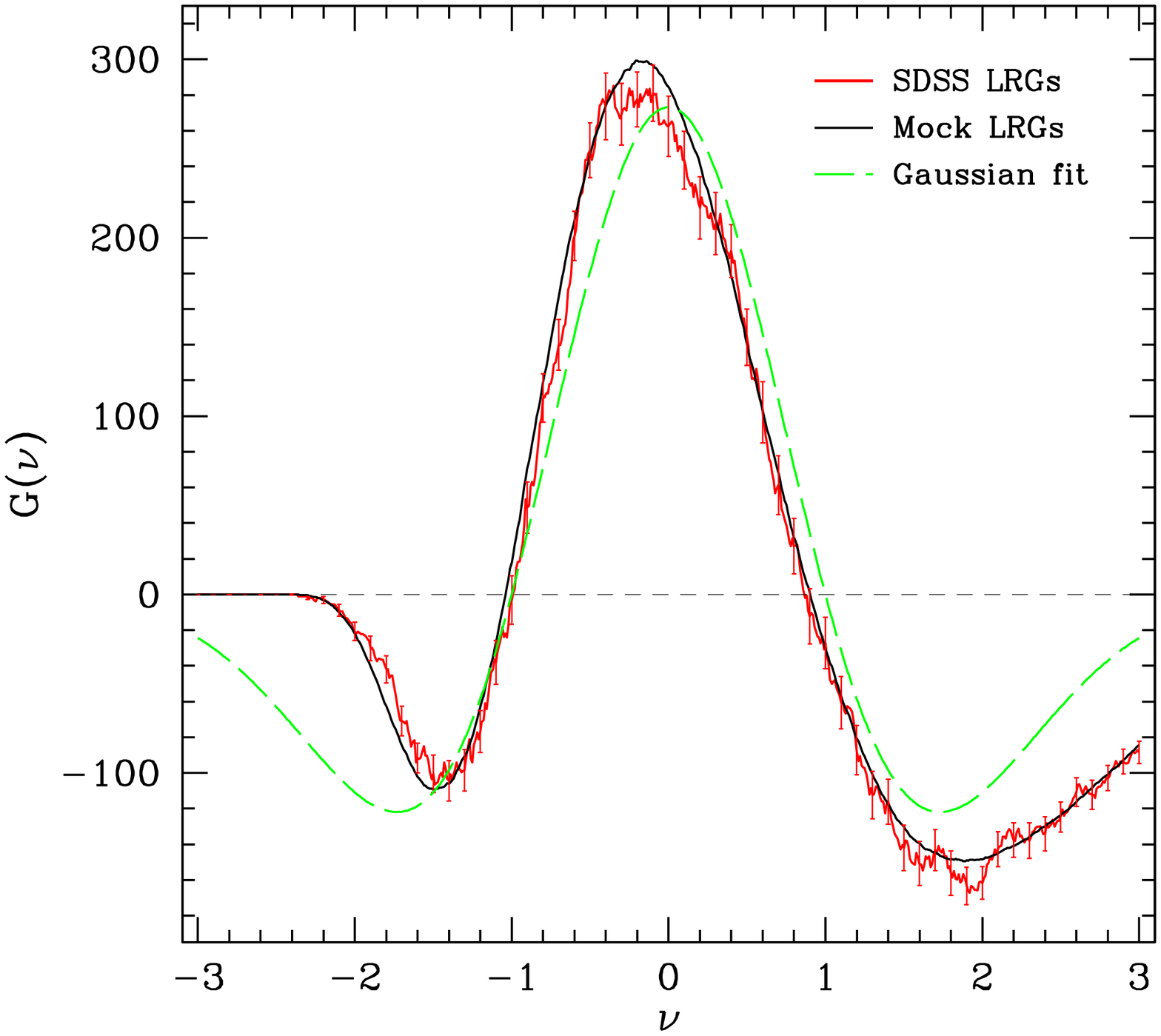}
\caption{Genus curve measured from the number density field
smoothed with $R_{\rm G} =22\hMpc$, from the SDSS sample
(red solid line, with error bars) and from mock samples (black solid line),
as a function of the direct density threshold $\nu$. The dashed line is
the best-fit Gaussian genus curve to the observed one.
}
\label{fig:gplot_dth}
\end{figure}

\begin{figure} 
\epsscale{0.8}
\plotone{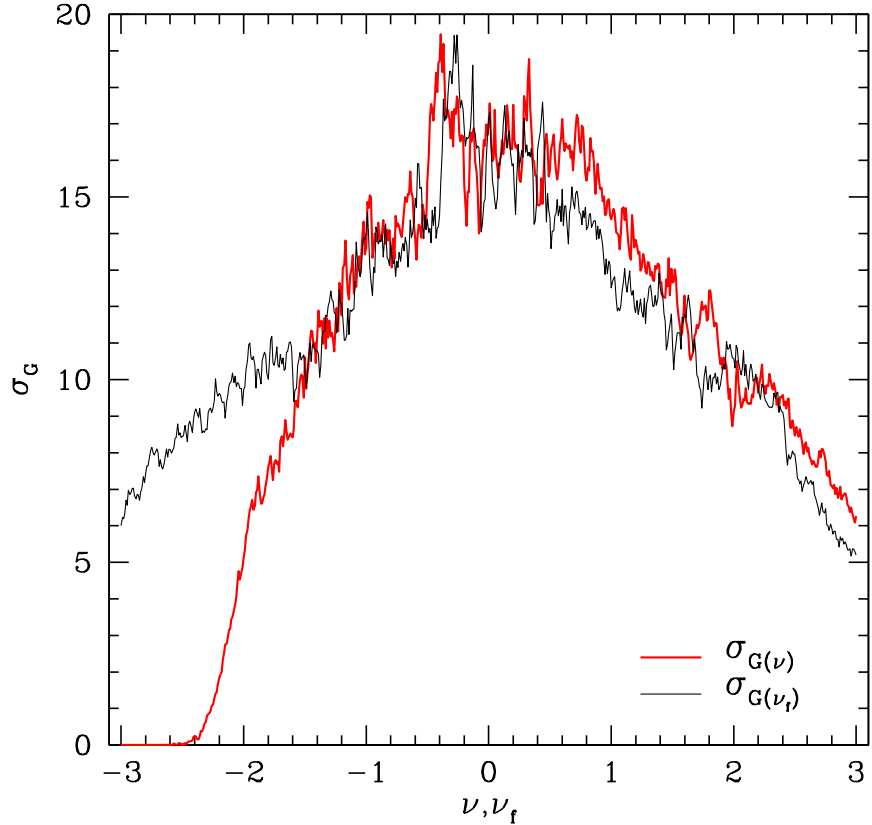}
\caption{Dispersion of the genus curve computed from the 81 mock samples,
as a function of the direct density threshold ($\nu$; red thick line)
and of the volume fraction ($\nu_{\rm f}$; black thin line). The dispersions at
threshold densities below $\nu\simeq -1.7$ suddenly decrease.
}
\label{fig:nuerr}
\end{figure}

Figure~\ref{fig:gplot_dth} shows the SDSS and mock genus curves, plotted versus the
direct density threshold $\nu$. These curves deviate
from the Gaussian predictions more than those obtained 
using the volume-fraction $\nu_{\rm f}$  shown in Figure~\ref{fig:main_result}.
Table~\ref{tab:mgplot_genus} from the previous Appendix~\ref{sec:genus_curves} lists
the genus values as a function of $\nu$.
By inspecting those values, we conclude that the direct density threshold parametrization 
is more sensitive to the skewness 
in the density probability distribution. However, given the vanishing 
genus and its large dispersion at threshold levels below
$\nu\simeq -1.7$ (see Figure~\ref{fig:nuerr}),
it is not appropriate to use this genus curve to
inspect the non-Gaussian deviation of the observational sample from
perturbation theory, at the smoothing scale adopted.

\end{document}